\documentclass[pdflatex,sn-mathphys-num]{sn-jnl}
\usepackage{graphicx}
\usepackage{multirow}
\usepackage{amsmath,amssymb,amsfonts}
\usepackage{amsthm}
\usepackage{mathrsfs}
\usepackage[title]{appendix}
\usepackage{xcolor}
\usepackage{textcomp}
\usepackage{manyfoot}
\usepackage{booktabs}
\newcommand{\uat}[2]{#1}
\usepackage{algorithm}
\usepackage{algorithmicx}
\usepackage{algpseudocode}
\usepackage{listings}

\theoremstyle{thmstyleone}

\theoremstyle{thmstyletwo}

\theoremstyle{thmstylethree}

\newcommand{\parder}[2]{ \frac{\partial #1}{\partial #2} }
\def \feix {Fe\,{\sc ix}}
\def \fexv {Fe\,{\sc xv}}
\def \fexix {Fe\,{\sc xix}}

\def \mgii {Mg\,{\sc ii}}
\def \caii {Ca\,{\sc ii}}
\def \hi {H\,{\sc i}}

\begin{document}

\title[Article Title]{On the ``cool" nature of coronal nanojets}

\author*[1]{\fnm{Gabriele} \sur{Cozzo}}\email{gabriele.cozzo@cfa.harvard.edu}

\author[1]{{\fnm{Paola} \sur{Testa}}\email{ptesta@cfa.harvard.edu}}

\author[2,3,4,5]{\fnm{Juan} \sur{Martinez-Sykora}\email{jmsykora@seti.org}}

\author[6,7]{\fnm{Paolo} \sur{Pagano}\email{paolo.pagano@unipa.it}}

\author[6,7]{\fnm{Fabio} \sur{Reale}\email{fabio.reale@unipa.it}}

\author[2,3,4]{\fnm{Bart} \sur{De Pontieu}}\email{bdp@lmsal.com}

\author[2,3,4,5]{\fnm{Viggo} \sur{Hansteen}}\email{vhansteen@seti.org}

\author[6]{\fnm{Alberto} \sur{Sainz-Dalda}\email{asainz@lmsal.com}}

\affil[1]{Harvard-Smithsonian Center for Astrophysics, 60 Garden St., Cambridge, MA 02193, USA}

\affil[2]{Lockheed Martin Solar and Astrophysics Laboratory, 3251 Hanover St, Palo Alto, CA 94304, USA}

\affil[3]{Rosseland Centre for Solar Physics, University of Oslo, P.O. Box 1029 Blindern, N-0315 Oslo, Norway}

\affil[4]{Institute of Theoretical Astrophysics, University of Oslo, P.O. Box 1029 Blindern, N-0315 Oslo, Norway}

\affil[5]{SETI Institute, 339 Bernardo Ave, Suite 200, Mountain View, CA, 94043, United States}

\affil[6]{INAF-Osservatorio Astronomico di Palermo, Piazza del Parlamento 1, I-90134 Palermo, Italy}

\affil[7]{Dipartimento di Fisica \& Chimica, Universit\`a di Palermo, Via Archirafi 36, 90123 Palermo, Italy}

\abstract{
Reconnection-driven nanoflares are widely considered a leading mechanism for coronal-loop heating, but their direct fingerprints in the tenuous coronal plasma remain elusive. The recently discovered coronal nanojets offer a potential probe of reconnection dynamics, but their extreme collimation, directionality and multi-wavelength visibility are not fully understood. Here we present a 3D rMHD simulation that unprecedentedly reproduces the key properties of nanojets, offering a viable model to explain their nature. These results provide a unified picture in which nanojet morphology, dynamics and detectability are contingent on the thermodynamic environment of reconnection.
Together, our results point to a cool origin of coronal nanojets, where cool and dense material permits narrow, multi-band jet signatures to emerge from reconnection.
}


\maketitle

\section{Introduction}
\label{introduction}

The bright solar corona consists of high temperature plasma largely confined in magnetic arches (coronal loops, \citep{reale2014coronal}), best visible in the EUV and X-ray bands. In active regions, typical temperatures are around 2 - 3 MK, and densities around $10^9$ cm$^{-3}$. Small reconnection events, referred to as  nanoflares \citep{parker1988nanoflares}, can, in principle, sustain such high temperature regimes, although observing the direct imprint of such time- and space-localised energy deposition remains a challenge.
Recent studies \citep{antolin2021reconnection} have shifted attention toward the dynamical aspects of reconnection, which always involves the impulsive release of magnetic tension and, therefore, plasma acceleration. In particular, a new class of small-scale jets was directly observed, and linked to reconnection-driven nanoflares.

Such ``nanojets'' \citep{antolin2021reconnection} are observed as narrow ($\lesssim 1\,\mathrm{Mm}$ width), short lived (tens of seconds), and fast (hundreds of $\mathrm{km}\,\mathrm{s}^{-1}$) jet-like bursts that travel perpendicularly to the guide field of coronal loops. Nanojets are detected in UV and often EUV images, forming a coherent multi-wavelength scenario \citep{antolin2021reconnection}. They are regarded as a likely direct signature of coronal heating by nanoflares \citep{parker1988nanoflares}, with magnetic reconnection proposed as their driving mechanism.

Originally detected \citep{antolin2021reconnection} with the Interface Region Imaging Spectrograph \citep[IRIS][]{de2014interface}, to date, observations from the Atmospheric Imaging Assembly (AIA) on board the Solar Dynamic Observatory (SDO) \cite{boerner2012initial, lemen2012atmospheric}, the Extreme Ultraviolet Imager (EUI) on board Solar Orbiter (SolO) \cite{muller2020solar,rochus2020solar}, and IRIS show nanojets primarily in the presence of clumpy (0.2" - 0.8" width), cold ($10^4 - 10^5 \mathrm{K}$) , and dense ($10^{10} - 10^{11}$ $\mathrm{cm^{-3}}$) 
plasma, such as coronal rain \citep{antolin2015multi, sukarmadji2022observations, sukarmadji2024transverse}, filaments \citep{liu2025deciphering}, and prominences \citep{gao2025reconnection, bura2025dynamics, wallace2025reconnection}. 
\textcolor{black}{A distinctive characteristic of these phenomena is their frequent occurrence as single-sided jets, with emission predominantly extending in one direction across the underlying magnetic structure, while clearly identifiable bidirectional counterparts are only rarely detected \citep{antolin2021reconnection}. Furthermore, nanojets are observed in a broad range of temperature-sensitive EUV passbands, implying a complex thermodynamic structure that likely encompasses both hot coronal plasma and cooler, denser components \citep{sukarmadji2022observations}. Although they are associated with small-scale energy release processes, such as nanoflares, nanojets are not observed ubiquitously throughout the solar corona, even in regions where such heating mechanisms are expected to operate pervasively.}

Small-angle magnetic reconnection has been invoked to explain many of their observed properties \citep{antolin2021reconnection}, including their bursty nature, as the rapid displacement of newly reconnected magnetic strands forces local plasma to be propelled perpendicularly to the loop's guide magnetic field. However, several key questions remain open: (I) How can their extreme collimation be reconciled with the global reconfiguration of reconnected field lines and the high thermal conductivity expected in the corona? (II) Why are nanojets most often observed as ``single-sided'' jets? (III) What physical conditions enable their multi-passband visibility? (IV) Why are nanojets not broadly observed everywhere in the corona where nanoflares are presumed to be important for its heating?

Here we show that narrow jets similar to observed nanojets are naturally produced when the magnetic field reconnects in an ambient  dense, relatively cool plasma. We also show that it is  necessary to distinguish  between ``reconnection outflows'', naturally arising as a dynamical outcome of small-angle magnetic reconnection, and ``nanojets'', which are also observational manifestations of the same process but that occur only under specific thermodynamic and geometric conditions.

Small-angle reconnection \citep{hesse1988theoretical, schindler1988general} can generate fast (sub-Alfvénic) outflows directed perpendicular to the guide field. These motions are driven by the rapid post-reconnection relaxation of magnetic tension. Plasma frozen into the field near the reconnection site is expelled from the current sheet as the newly reconnected field lines retract. In coronal loops, this process is generally expected to involve plasma along the entire reconnected field line, as the retraction affects the full flux tube rather than a limited portion of it \citep{antolin2021reconnection, 2025A&A...695A..40C}.

Both 2.5D and fully 3D numerical experiments \citep{antolin2021reconnection, pagano2021modelling, 2025A&A...695A..40C, sen2025merging} reproduced this mechanism, showing that small-angle reconnection can (I) convert nanoflare-scale magnetic energy into heat; (II) raise plasma temperatures to several million degrees Kelvin; and (III) accelerate fast outflows perpendicular to the guide field. It was further shown \cite{Cozzo_2026a,Cozzo_2026b} that reconnection outflows can be common and frequent, accompanying heating by magnetic energy release in numerous and widespread current sheets. Reconnection outflows can therefore be regarded as the dynamical aftermath of Parker’s topological dissipation \citep{parker1972topological}, in which tangential discontinuities in braided fields collapse into current sheets \citep{rappazzo2013current} and reconnect. If these events occur in the still unheated and therefore tenuous corona, they are likely difficult to detect with current EUV imagers \citep{2025A&A...695A..40C}.

In this paper, we propose that the observed narrow nanojets correspond to “bullets” of cold, dense plasma that are dragged by rapidly evolving, reconnected magnetic field lines. We present a self-consistent scenario for nanojet formation that accounts for (I) the observed morphology, including sub-Mm collimation across the guide-field direction (as well as across the field) and predominant unidirectionality; (II) multi-wavelength visibility, including IRIS 1400~\AA\ and the optically thin SDO/AIA channels; and (III) a reconnection-driven energy release at nanoflare scales, accompanied by high-speed flows perpendicular to the magnetic field.

We report a fully 3D MHD simulation (with the PLUTO code \cite{mignone2007pluto}) of a kink-unstable solar filament that addresses these key points. We will focus on a reference nanojet case. A second case is reported in Appendix \ref{sec:secondnanojet}.

\section{Results}
\label{sec:results}

\begin{figure}[t]
\centering
\includegraphics[width=\hsize]{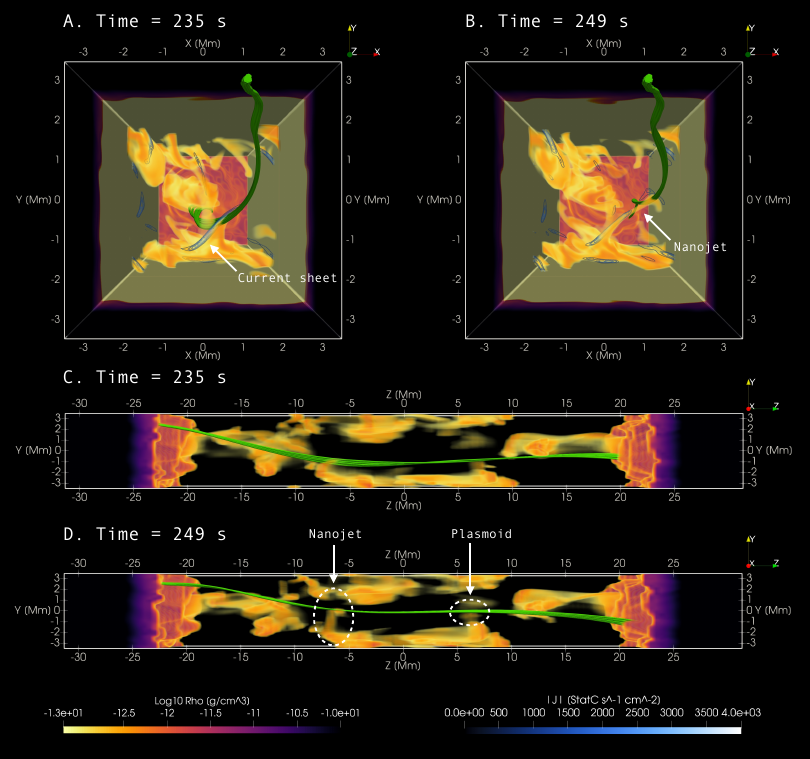}
\caption{3D view of the nanojet. Volume rendering of the simulation box before ($t = 234\,\mathrm{s}$, panels A and C) and during ($t = 249\,\mathrm{s}$, panels B and D) the nanojet evolution and from two different perspectives (the $\hat z$ direction is parallel to the guide field). The volume rendering shows the plasma density. A reconnecting bundle of field lines (green) is traced from the z-negative footpoint. In panels A and B blue contours locate the current sheets. A supplementary video is available.}
\label{fig:3D_rendering}
\end{figure}

\begin{figure*}[t]
\centering
\includegraphics[width=\hsize]{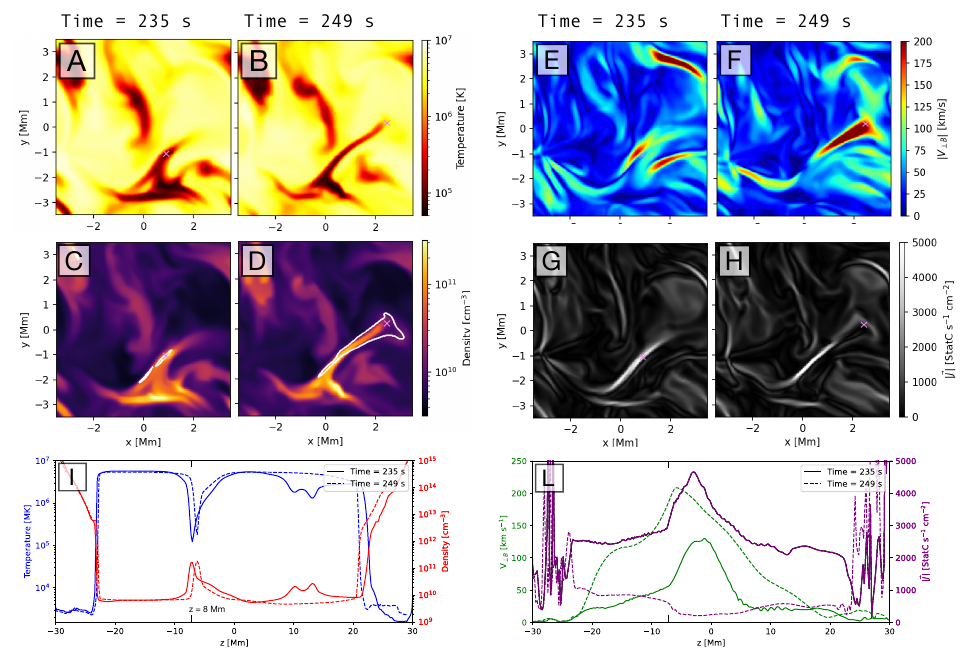}
\caption{Physical properties of the reference nanojet. Panels A to H show a horizontal slice ($z = 8\,\mathrm{Mm}$) of the simulation box before (t = 234 s) and during (t = 249 s) the nanojet evolution, and include maps of the temperature (A and B), plasma density (C and D), magnitude of velocity perpendicular to the guide magnetic field (E and F), and current density (G and H). Panels I and L show the stratification of the same quantities along a field line passing through the nanojet ($\times$ symbol in panels A-H). In panels C and D the white contour identifies the reconnection outflow as detected by the ROAD algorithm \citep{Cozzo_2026b}. A supplementary video is available.}
\label{fig:horizontal_slice}
\end{figure*}

\begin{figure*}[t]
\centering
\includegraphics[width=\hsize]{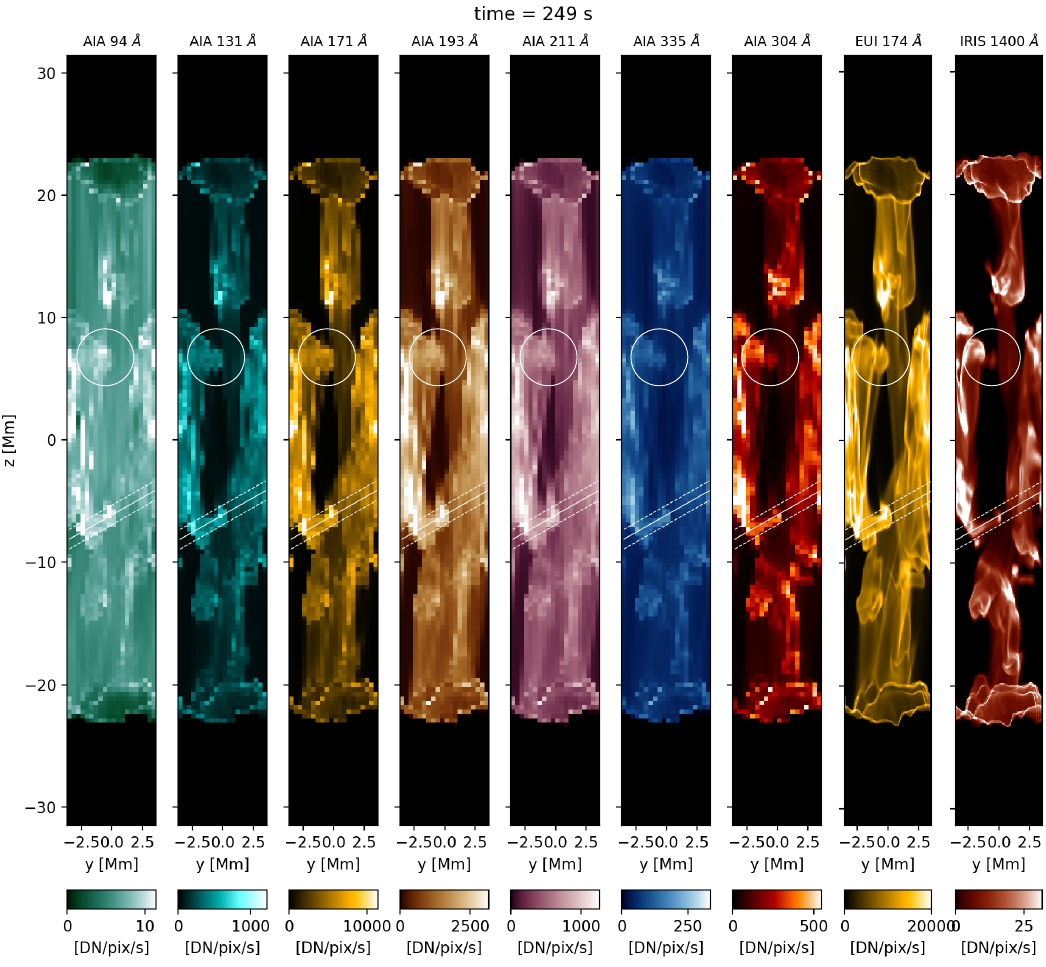}
\caption{Synthesis of the EUV emission across SDO/AIA pass bands at 94 \AA, 131 \AA, 171 \AA, 193 \AA, 211 \AA, 335 \AA, and 304 \AA, SolO/EUI channel at 174 \AA, and IRIS SJI channel at 1400 \AA, respectively. A solid white line aligns with the nanojet direction of propagation, dashed lines visually estimate the width. A circle is placed around a plasmoid. A supplementary video is available.}
\label{fig:forward_modelling_n1}
\end{figure*}

\begin{figure*}[t]
\centering
\includegraphics[width=\hsize]{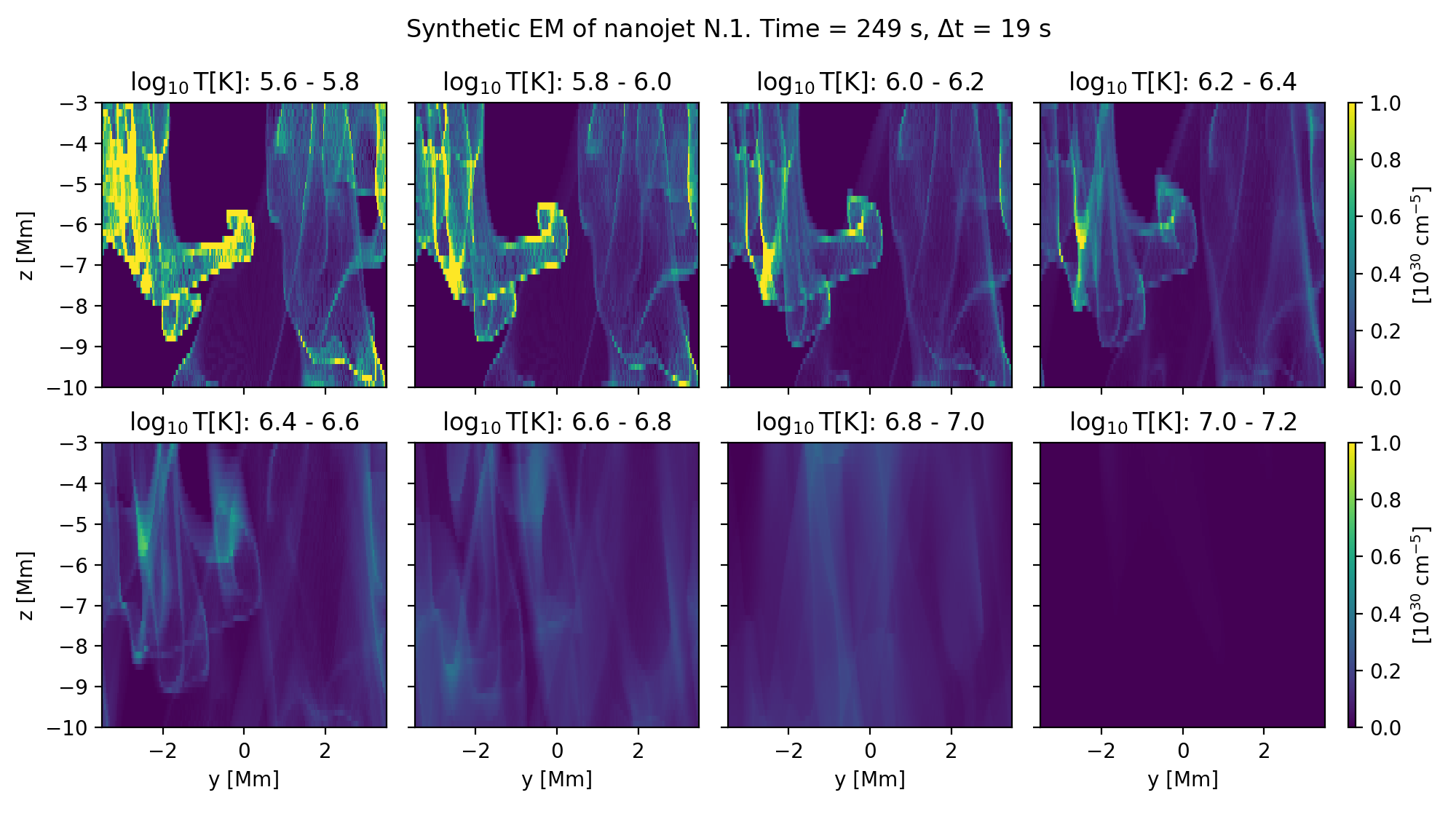}
\caption{Nanojet Emission Measure (EM), as extracted from the simulation data, across 8 temperature bins in the range $5.6 < \log T < 7.2$ (the same as in Fig. \ref{fig:DEM_inversion_n1}). A supplementary video is available.}
\label{fig:DEM_synthesis_n1}
\end{figure*}
\backmatter

We model a scenario similar to that observed by \citep{bura2025dynamics}, namely a confined flare associated with a non-eruptive filament. \textcolor{black}{The event, occurred on October 13, 2023, developed from a system of braided and twisted coronal loops into a C-class flare with no large-scale magnetic field restructuring or filament eruption, but with substantial evidence of reconnection driven nanojets. In our simulation, we adopted the initial condition of a twisted magnetic flux tube with} an average field strength of $80,\mathrm{G}$, containing cold, dense material trapped within the tube (Fig.~\ref{fig:initial_conditions} in sec \ref{sec:numericalmodel}). Energy release is triggered by the onset of a kink instability, which disrupts the mass condensation inside the flux rope (Fig.~\ref{fig:kink_instability} and further discussion in appendix \ref{sec:kinkintability}).

Figure~\ref{fig:3D_rendering} (A, C) shows the system at $t=235\,\mathrm{s}$ after the kink instability begins. The original filament has fragmented and is dispersed throughout the domain. A current sheet forms within a cloud of dense plasma, and a bundle of field lines threads the sheet. Within a few seconds (Fig.~\ref{fig:3D_rendering} B, D), the field lines reconnect and rapidly retract away from the current sheet, dragging a narrow plasma jet.

Although reconnection deposits thermal energy, the cold, dense jet plasma undergoes only limited heating because of its relatively large heat capacity; most of the temperature increase occurs in the surrounding tenuous plasma (Fig.~\ref{fig:3D_rendering_A1}). 
The reconnection point ($z\simeq 0\,\mathrm{Mm}$, \textcolor{black}{as found from field lines evolution, Fig.~\ref{fig:3D_rendering_A2}}) is about $7\,\mathrm{Mm}$ away from the jet. Of the two reconnecting flux systems (see Fig.~\ref{fig:3D_rendering_A2}  and refer to the discussion in Appendix \ref{sec:reconnectiondynamics}), only the bundle shown in Fig.~\ref{fig:3D_rendering} encounters cold, dense plasma and consequently accelerates a nanojet. A more tenuous plasmoid is also ejected around $z = 6\,\mathrm{Mm}$ (Fig.~\ref{fig:3D_rendering} D).

These results indicate that nanojets can also develop away from the reconnection site and be accelerated primarily by the dragging of newly reconnected field lines rather than by the local retraction of magnetic field lines at the current sheet. \textcolor{black}{This implies that the canonical picture of a rapid (possibly shock-like) retraction of magnetic field in the immediate vicinity of the reconnection region may not be required for nanojet observability. Moreover, canonical interpretation predicts bidirectional jets, a signature that is seldom observed, as most of the events appear predominantly single-sided. In contrast, the scenario presented here naturally reproduces this asymmetry, offering a more robust interpretation of nanojet observations.}
It also provides an alternative interpretation for clusters of nanojets \citep{antolin2021reconnection}: they likely reflect a single reconnection episode interacting with clumpy, dense plasma distributed along the reconnected field lines, rather than multiple reconnection events closely spaced in space and time.

Figure~\ref{fig:horizontal_slice} shows the plasma properties in a cross-field cut through the loop at the jet location (top panels) and along the magnetic field line shown in Fig.~\ref{fig:3D_rendering} (bottom panels). Panels A–D and I show that the nanojet plasma remains cold and dense, with transition-region temperatures (minimum values just above $10^5\,\mathrm{K}$) and densities roughly an order of magnitude above the background, locally reaching $10^{11}\,\mathrm{cm}^{-3}$. Panels E–H and L show that the event simultaneously exhibits the dynamical signatures of reconnection outflows, as reported in previous modelling studies (e.g. \cite{antolin2021reconnection}; \cite{2025A&A...695A..40C}; \cite{Cozzo_2026b}): fast tails of plasma moving at a few hundred $\mathrm{km}\,\mathrm{s}^{-1}$ predominantly perpendicular to the magnetic field, in association with a rapidly dissipating current sheet. Specifically, the temperature drop and density enhancement are confined to a narrow segment of the field line, whereas the velocity enhancement and current-density signature extend for several Mm along the guide-field direction. This separation highlights the distinction between reconnection outflows (defined by $\vec{V}_{\perp B}$ and $\vec{J}$, \cite{Cozzo_2026b}) and nanojets, which require specific thermodynamic conditions (dense, cold plasma embedded within the outflow).

Forward modelling confirms that the same event is detectable across multiple passbands. Fig. \ref{fig:forward_modelling_n1} shows synthetic EUV emission from the flux tube: the nanojet appears as a short-lived, narrow jet-like feature in all SDO/AIA optically thin EUV channels, in the 174~\AA\ channel of SolO/EUI, and in the IRIS 1400~\AA\ channel, consistent with the multi-wavelength scenario reported observationally \citep{antolin2021reconnection, sukarmadji2022observations}. \textcolor{black}{At the same snapshot time,} a moving plasmoid is also visible at $z \sim 6\,\mathrm{Mm}$ (white circle).

The nanojet propagates for $\sim 30\,\mathrm{s}$ and reaches a maximum plane (LoS) projected length of $\sim 2\,\mathrm{Mm}$.

The nanojet propagates for $\sim 30,\mathrm{s}$ and reaches a maximum projected length of $\sim 2,\mathrm{Mm}$ in the plane of the sky. From the time–distance maps (Fig.~\ref{fig:time_distance_maps_1}), the mean projected in-plane speed is $\sim 70\,\mathrm{km}\,\mathrm{s}^{-1}$. \textcolor{black}{LoS velocities will be measurable with high cadence coronal spectrographs, such as the forthcoming Multislit Solar Explorer (MUSE \citep{de2020multi, de2022probing}),} enabling an estimate of the field-line propagation speed by combining their plane-of-sky and spectroscopic components (Fig.~\ref{fig:MUSE_FM_n1}).

The apparent jet width depends on the diagnostic, \textcolor{black}{not dissimilarly to reported widths of coronal rain clumps \citep{antolin2015multi}}. In AIA 304~\AA\ and IRIS 1400~\AA, the jet appears narrower (width $\lesssim 1\,\mathrm{Mm}$) because the emission is dominated by a thin core at TR temperatures ($\sim 10^5\,\mathrm{K}$, Fig. \ref{fig:emission_profile}). In hotter AIA channels, emission from a surrounding envelope, where the temperature transition from transition-region to multi-million-Kelvin coronal values broadens the apparent structure (width $\gtrsim 1\,\mathrm{Mm}$, see EM analysis in Fig.~\ref{fig:DEM_synthesis_n1}). Nevertheless, the dominant contribution to the signal across the channels originates from the \textcolor{black}{cool TR} component, as shown by the emission-measure analysis (Fig.~\ref{fig:DEM_synthesis_n1}).

Finally, we caution that EM inversions from AIA data can produce misleading high-temperature components when performed over the standard range $\log_{10}T\, [\mathrm{K}]\in(5.6,7.2)$, because the multi-peaked AIA temperature response functions allow dense, cool plasma to mimic hot emission \textcolor{black}{\citep{peter2012catastrophic}}. This effect is evident in Fig.~\ref{fig:DEM_inversion_n1} (e.g. the $\log_{10}T\,[\mathrm{K}]=6.4$–$6.6$ bin, to be compared with Fig.~\ref{fig:DEM_synthesis_n1}) and can lead to an overestimated jet temperature (Fig.~\ref{fig:DEM_weigthted_temperature_n1}).

\section{Discussion}
\label{sec:discussion}

\begin{figure*}[t]
\centering
\includegraphics[width=\hsize]{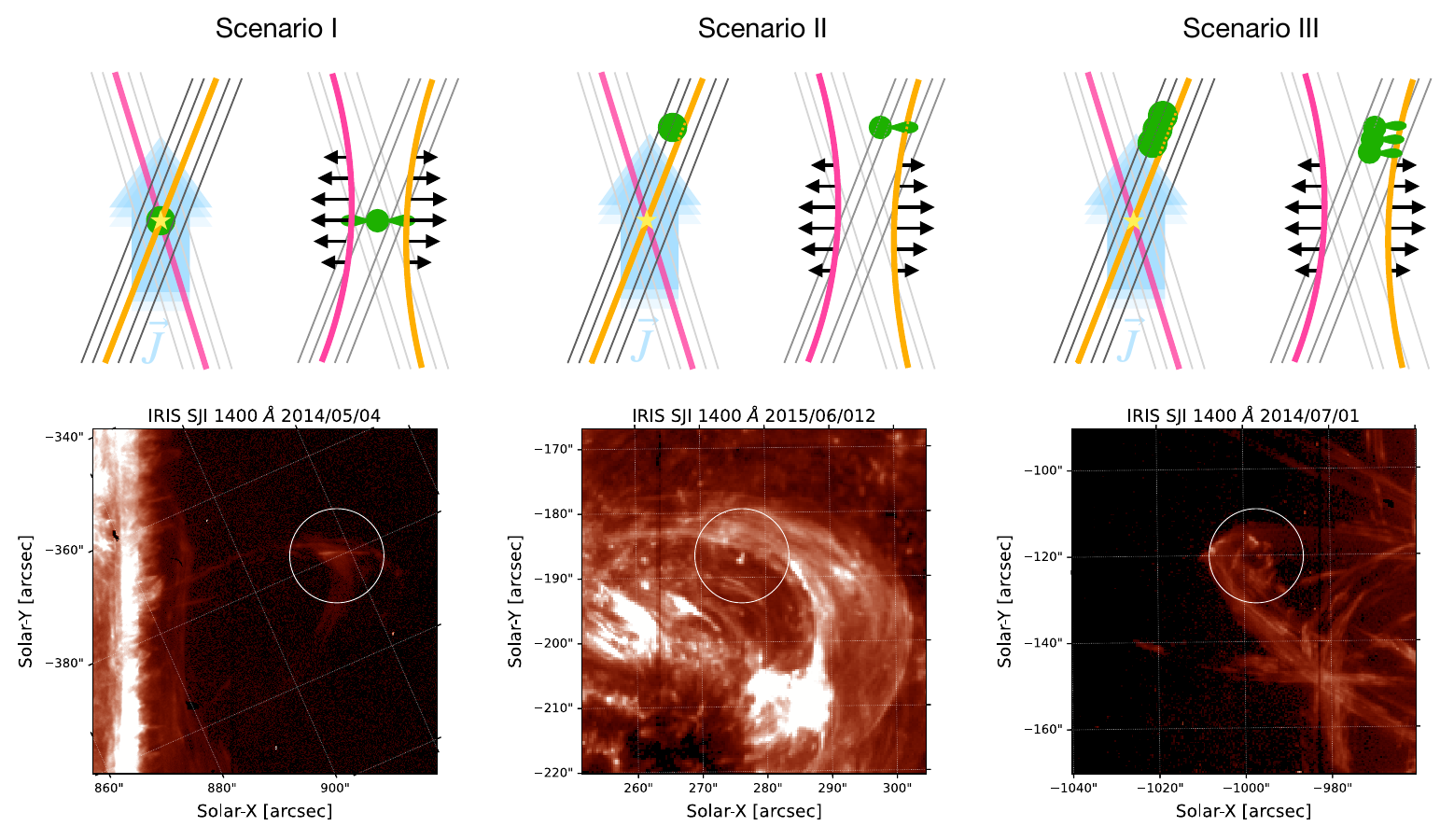}
\caption{Three different scenarios leading to different types of nanojets: bidirectional (left), unidirectional (middle), or cluster (right), depending on the location of the dense plasma blob (green circle) relative to the reconnection point (yellow star). \textcolor{black}{Observational examples in the IRIS SJI at 1400 \AA\ channel are provided for each case in the lower panels (see \citep{sukarmadji2022observations} for observations details).}}
\label{fig:Scenarios}
\end{figure*}
\backmatter
 
The extreme collimation of coronal nanojets is difficult to reconcile with component magnetic reconnection alone: in the presence of a strong guide field, the reconnection-driven outflow involves the retraction of field lines over a large spatial scale, larger than the observed width of a nanojet. In addition, the efficient field-aligned thermal conduction is expected to smooth temperature gradients along the reconnected flux tube. Together, these effects tend to broaden the thermodynamic signatures of reconnection outflows along the guide-field direction, in apparent contrast to the observed narrow morphology of nanojets.

Another outstanding issue concerns their directionality, since magnetic reconnection naturally produces bidirectional outflows. Although loop curvature can attenuate one branch of the flow \citep{pagano2021modelling}, 
observations show instances of inward- as well as outward-propagating jets across curved structures \citep{patel2022hi}, suggesting that the preferential propagation direction is not determined solely by the large-scale loop geometry. Overall, existing models have successfully linked nanojet-like dynamics to reconnection, but a unified framework that simultaneously explains their morphology, directionality, and multi-band detectability is still missing. For example, the 3D MHD-avalanche model of \citep{2025A&A...695A..40C} reproduces jet-like emission in the hot MUSE Fe~XIX line, but does not produce detectable counterparts in AIA passbands. The 2.5D coalescing–flux-rope model of \citep{sen2026extreme} represents an important step forward, yielding outflows visible in some AIA channels, but does not account for the full 3D dynamics of reconnection and nanojet evolution.

Our results suggest that the missing ingredient does not concern the reconnection mechanism itself, but rather the environment in which reconnection outflows develop. Specifically, the presence and spatial distribution of cold, dense plasma along the reconnecting field can determine whether a reconnection outflow becomes observable as a nanojet, and whether it appears single-sided or double-sided. This leads to three possible scenarios: (I) if cold plasma is located at the reconnection site, a double-sided nanojet can be produced as both outflow branches involve dense material (Fig. \ref{fig:Scenarios}, left); (II) if only one set of reconnected field lines intersects a localised “bubble” of cold plasma, the event will produce a single-sided nanojet, with the opposite branch remaining tenuous and comparatively faint (Fig. \ref{fig:Scenarios}, center); (III) if the reconnecting field lines are embedded within a clumpy cloud of dense material, a cluster of nanojets can emerge from a single reconnection episode, as the field lines bundle crosses distinct condensations (Fig. \ref{fig:Scenarios}, right).

Scenarios (II) and (III) are consistent with observations in which single-sided jets (either isolated or clustered) originate from the edges of coronal loops or loop strands seen in the IRIS 1400~\AA\ channel \citep{antolin2021reconnection, sukarmadji2022observations}. By contrast, double-sided jets, scenario (I), tend to originate from within the loop volume \citep{sukarmadji2022observations, sukarmadji2024transverse}. \textcolor{black}{Such a scenario is less probable than Scenario II, as it relies on a highly constrained spatial distribution of the cold plasma, which may account for its limited observational occurrence.} \citet{sukarmadji2022observations} also report events in which a loop appears to split into two strands, which may represent a limiting case where reconnected field lines interact with dense plasma over a substantial fraction of the loop length.

The reference nanojet turns into a floating plasmoid after the initial journey across the guide magnetic field (Fig. \ref{fig:plasmoid}): \textcolor{black}{it gradually loses its collimated, jet-like structure and evolves into a more compact, clump-like morphology.} Observations of plasmoids resulting from nanojets traveling along the new field line \citep{sukarmadji2022observations} further support this scenario.

Finally, our results help clarify the origin of the reported multi-wavelength signatures of nanojets. While nanojets can span a range of temperatures, from a cold transition-region core to a warmer coronal envelope \citep{antolin2021reconnection, sukarmadji2022observations}, their multi-wavelength observability may still be dominated by the cold, dense component, which carries most of the emission measure. In particular, the broad temperature response functions of several AIA channels make them sensitive to dense plasma at relatively low temperatures, which can produce apparent hot coronal emission even when the bulk of their volume remains cold.

\section{Conclusions}
\label{sec:conclusions}

In this work, we bridge the gap between reconnection outflows and the observed phenomenon of coronal nanojets. Using a fully 3D MHD simulation of a kink-unstable, filament-hosting flux tube, we show that small-angle reconnection can accelerate narrow, short-lived jets when, and only when, the newly reconnected field lines encounter cold, dense plasma. In this regime, the nanojet originates from a dense “bullet” accelerated by rapidly relaxing field lines; as a result, the observable nanojet can form away from the reconnection point, appear preferentially single-sided, and remain strongly collimated despite the global field-line reconfiguration, \textcolor{black}{given the inefficient thermal conduction at low TR temperatures}. Forward modeling further demonstrates that the same cold component can dominate the emission measure while still producing signatures across multiple wavelengths, naturally accounting for the reported multi-band visibility. Together, these results provide a unified physical picture in which the morphology, directionality and multi-wavelength detectability of nanojets are controlled primarily by cold/dense material rather than by reconnection geometry alone.

\textcolor{black}{The requirement of low-temperature plasma in these events points to a potentially important role of non-ideal effects beyond single-fluid MHD. In particular, Pedersen dissipation and ion–neutral coupling may play a role in triggering a certain class of nanojets. This suggest that nanojets may have multi-physical processes acting across different regimes. Establishing this connection will require dedicated multifluid and multispecies modeling \citep[e.g.,][]{martinez2020velocity}.}

\section{Methods}
\label{sec:methods}

\subsection{Numerical model}
\label{sec:numericalmodel}
\begin{figure*}[t]
\centering
\includegraphics[width=\hsize]{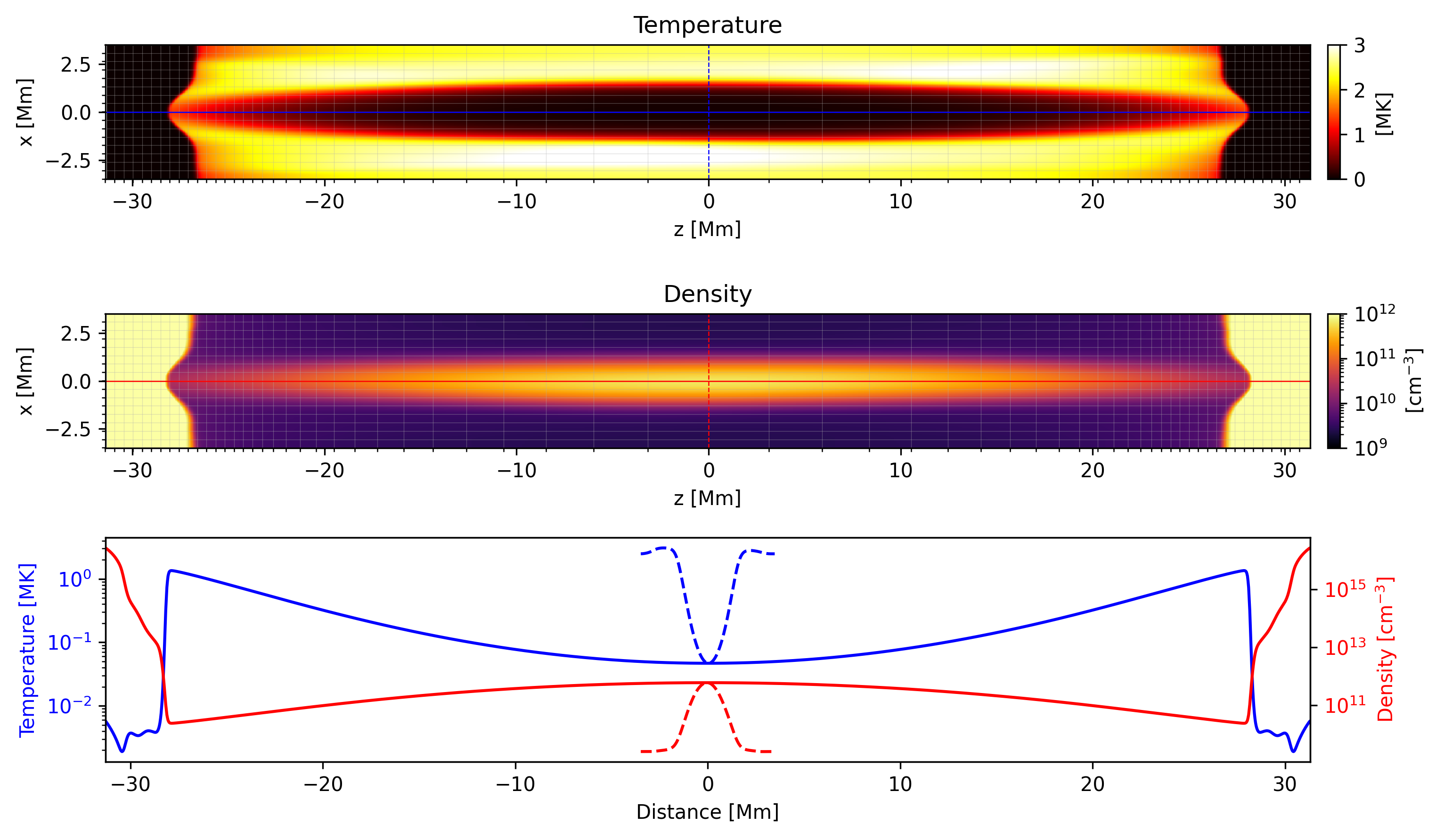}
\caption{Initial conditions of simulated the solar atmosphere. Top panel: temperature map across the box (y = 0 Mm plane). Mid panel: Density map across the box. Bottom panel" plot of the temperature (red) and density (blue) along z (solid) and x (dashed) axes.}
\label{fig:initial_conditions}
\end{figure*}
\backmatter

This work is based on a 3D MHD model of a magnetized solar atmosphere in which a twisted coronal flux rope, hosting a cool ($\sim 10^5\,\mathrm{K}$), dense filament ($10^{11}\,\mathrm{cm}^{-3}$), undergoes a kink instability and fragments into sub-arcsecond strands.

The simulations are performed using the PLUTO code \citep{mignone2007pluto}, a modular Godunov-type code for astrophysical plasmas. The evolution of the plasma and magnetic field is obtained by solving the full, time-dependent MHD equations for a resistive, fully ionized plasma, including gravity, thermal conduction \citep{spitzer1953transport}, radiative losses \citep{Carlsson2012}, and an anomalous magnetic diffusivity \citep{hood2009coronal}:
\begin{align}
&\parder{\rho}{t} + \mathbf{\nabla} \cdot (\rho \vec v) = 0, \label{Eq:PLUTO_MHD_1}\\
&\parder{\rho \vec v}{t} + \mathbf{\nabla} \cdot (\rho \vec v \vec v) = - \nabla \cdot (\mathbf{I} P + \mathbf{I} \frac{B^2}{8 \pi} - \frac{\vec B \vec B}{8 \pi}) + \rho \vec g, \label{Eq:PLUTO_MHD_2}\\
&\parder{B}{t} - \nabla \times (\vec v \times \vec B) = - \eta \nabla^2 \vec B, \label{Eq:PLUTO_MHD_3}\\
&\parder{}{t} \left( \frac{B^2}{8 \pi} + \frac{1}{2} \rho v^2 + \rho \epsilon + \rho g h\right) + \nabla \cdot [ \frac{c}{4 \pi} \vec E \times \vec B + \frac{1}{2}\rho v^2 \vec v \\ & + \frac{\gamma}{\gamma-1} P \vec v + \vec F_{\mathrm{c}} + \rho g h \vec v ] = Q(T), \nonumber \label{Eq:PLUTO_MHD_4} \\
& P = (\gamma -1) \rho \epsilon = \frac{2 k_B}{\mu m_H} \rho T, 
\end{align}
where $t$ is the time; 
$\rho$ is the mass density; 
$\vec{v}$ is the plasma velocity; $P$ is the thermal pressure; 
$\vec{B}$ is the magnetic field; $\vec E$ is the electric field;
$\vec{g}$ 
is the gravity acceleration vector for a curved loop;  $\mathbf{I}$ is the
identity tensor;
$\epsilon$ is the internal energy; 
$\eta$ is the magnetic
diffusivity;  
$T$ is the temperature; 
$\vec{F}_{\mathrm{c}}$ is the thermal conductive flux; $Q(T)$ the radiated energy losses; $m_H$ is the hydrogen mass; $k_B$  is the Boltzmann constant;
$\mu = 1.265$ is the mean ionic weight (in units of proton mass, and assuming coronal elemental abundances) \cite{anders1989abundances}.

The differential equations are numerically integrated using the PLUTO MHD module \citep{mignone2007pluto}, configured to compute intercell fluxes with the Harten-Lax-Van Leer approximate Riemann solver \citep{roe1986characteristic}, while second-order accuracy in time is achieved using a Runge-Kutta scheme. A Van Leer limiter \citep{sweby1985high} for the primitive variables is used. 
The solenoidal condition is maintained with the constrained transport approach \citep{balsara1999staggered}.

We assumed the field-line-aligned gravity of a typical semicircular coronal loop:
\citep{reale20163d, cozzo2023coronal}:
\begin{equation}
g(z) \hat{\mathbf{z}} =
\begin{cases}
 g_{\odot} \sin{\left(\pi \frac{z}{L}\right)} \,\hat{\mathbf{z}}, & |z| \le 25\,\mathrm{Mm}, \\
 g_{\odot} \, \mathrm{Sign}\,(z) \,\hat{\mathbf{z}} & |z| > 25\,\mathrm{Mm},
\end{cases}
\end{equation}
where $ g_{\odot}  = \frac{G M_{\odot}}{R_{\odot}^2}$ is the gravitational acceleration at the solar surface.

The thermal conductive flux is defined as follows:
\begin{align}
& \vec{F}_c = \frac{F_{\mathrm{sat.}}}{F_{\mathrm{sat.}} + |F_{\mathrm{class.}}|} \vec{F}_{\mathrm{class.}}, \\
& \vec{F}_{\mathrm{class.}} = - k_{\parallel} \hat{\vec{b}} ( \hat{\vec{b}} \cdot \mathbf{\nabla} T) - k_{\perp} \left[ \mathbf{\nabla} T - \hat{\vec{b}} ( \hat{\vec{b}} \cdot \mathbf{\nabla} T) \right], \\
& F_{\mathrm{sat.}} = 5 \phi \rho c^3_{\mathrm{iso.}},
\end{align}
where, $k_{\parallel} = 9.2 \times 10^{-7} T^{\frac{5}{2}}$ and
$k_{\perp} = K_{\perp} = 5.4 \times 10^{-16} \rho^2 / (B^2T^{\frac{1}{2}})$, $c_{\mathrm{iso.}}$ is the isothermal sound speed; $\phi = 0.9$ is a dimensionless free parameter; $\hat{\vec{b}} = \vec B / B$ is the magnetic field unit vector; and $F_{\mathrm{sat.}}$ is the saturated flux.

We account for optically thin radiative cooling in the corona with a rate per unit volume:
\begin{equation}
    Q(T) = - \Lambda(T) n_e n_H,
\end{equation}
where $\Lambda (T)$ are the radiative rates per unit emission measure (from the CHIANTI v.~7.0 database \citep{landi2013prominence}, assuming coronal element abundances 
\citep{feldman1992elemental}), while $n_H$ ($n_e$) is the hydrogen (electron) number density. 

We also included an accurate description of chromospheric radiative losses assuming, in those conditions, radiative energy balance to be dominated by a small number of strong lines, including neutral hydrogen (\hi), singly ionized calcium (\caii), and 
singly ionized magnesium (\mgii) \citep{vernazza1981structure}.
We used a detailed radiative transfer calculations \cite{Carlsson2012} and the derived lookup tables to approximate the net effects of electron-impact excitation/radiative deexcitation processes as the product of a optically thin radiative loss function ($L_{X_m} (T)$, for the element $X$ in the ionisation stage $m$), an escape probability (for the photon to escape from the chromosphere, $E_{X_m} (\tau)$), and the element fraction of ionization ($N_{X_m}/N_X$):
\begin{equation}
    Q_X = - L_{X_m} (T) \, E_{X_m} (\tau) \, \frac{N_{X_m}}{N_X} (T) A_X \frac{N_H}{\rho} n_e \rho,
    \label{Eq:chromospheric_cooling}
\end{equation}
where $A_X$ is the abundance of the element $X$, and $N_H = 4.407 \times 10^{23}$ is the number of hydrogen ($H$) particles per gram of solar material \citep{Carlsson2012}. 

For an accurate description of the plasma flows across the transition region \citep{bradshaw2013influence} and the filament–corona transition regions \citep{johnston2025filament}, we adopted the Adaptive Conduction method \citep[TRAC;][]{johnston2019fast,johnston2020modelling,johnston2021fast}.

We considered an anomalous plasma resistivity \citep{hood2009coronal,reale20163d} that is switched on only in the corona and TR (i.e. above $T_{\mathrm{cr}} = 10^4\,\mathrm{K}$) where the magnitude of the current density exceeds a threshold:
\begin{equation}
\eta =
    \begin{cases}
    \eta_0 & J \ge J_{\mathrm{cr}}  \text{ and } T \ge T_{\mathrm{cr}} \\
    0  & J < J_{\mathrm{cr}} \text{ or } T < T_{\mathrm{cr}}
    \end{cases},
    \label{Eq:anomalus_diffusivity}
\end{equation}
where $\eta_0 = 10^{11}\,\mathrm{cm}^{2}\,\mathrm{s}^{-1}$  and $J_{\mathrm{cr}} = 1200\,\mathrm{StatC}\,\mathrm{cm}^{-2}\,\mathrm{s}^{-1}$ \citep{reale2025connection}. 

We consider a 3D Cartesian box of length $62\,\mathrm{Mm}$, aspect ratio $\sim 10$ ($-x_M < x < x_M$, $-y_M < y < y_M$, and $-z_M < z < z_M$, where $x_M = y_M = 7\,\mathrm{Mm}$ and $z_M = 31\,\mathrm{Mm}$). The equations are solved inside the box over a stratified, magnetized solar atmosphere, that includes a high-beta chromosphere and a tenuous coronal environment. The coronal volume is $50\,\mathrm{Mm}$ long and extends along $\hat z$. It is enclosed by two chromospheric regions, at the opposite sides of the box, from which it is separated by a thin interface, the transition region. Specifically, we adopted a staggered grid, uniform along $\hat x$ and $\hat y$, with $\Delta x = \Delta y \sim 30\,\mathrm{km}$. Along $\hat z$ we use a non-uniform grid, with high resolution ($\Delta z \sim 25 \,\mathrm{km}$) in the chromosphere and TR, decreasing logarithmically with height in the corona (up to $\Delta z \sim 100\,\mathrm{km}$). 
The magnetic field is modeled as a straight coronal flux tube connecting the two opposite sides of the box. Since the tube is much longer than wider, and except for the loop-aligned gravity, curvature effects are neglected. \textcolor{black}{In this case, asymmetries in nanojet propagation are not determined by the inward-directed magnetic tension characteristic of curved flux tubes, although certain regimes may exist where the outward-directed tension is too weak to produce a detectable outward jet, leading to an inward directed jet only.}

The flux tube is rooted in the chromospheric regions that act as two distant, independent footpoints and mass reservoirs for the corona. The magnetic field has a fairly uniform strength in the corona of 80 G and tapers in the chromosphere to reach a thousand Gauss \citep{guarrasi2014mhd}.

The boundary conditions (BCs) are periodic at $x = \pm x_M$ and $y = \pm y_M$. Reflecting BCs for the velocity and symmetric BCs for pressure and density were set at $z = \pm z_M$. The magnetic field is line-tied to the upper and lower boundaries (no field line slippage is allowed across the dense, highly conducting photosphere) and is subject to antisymmetric BCs. Magnetic footpoints move accordingly to prescribed boundary motions only. Specifically, rotational motions at the upper and lower footpoint boundaries twist the flux tubes around the central axis \citep{cozzo2023coronal, reale20163d} with a constant (rigid-body-like) angular velocity $\omega(r)$ that decreases linearly in an outer annulus: \citep{reale20163d}: 
\begin{equation}
    v_{\phi} = \omega \, r, 
\end{equation}
with
\begin{equation}
    \omega = \omega_0 \times
    \begin{cases}
    1 & r < r_{\mathrm{max}}   \\
    (2 r_{\mathrm{max}} - r)/r_{\mathrm{max}} & r_{\mathrm{max}} < r < 2 r_{\mathrm{max}}    \\
    0 & r > 2 r_{\mathrm{max}}
    \end{cases}
    \label{eq:angular_vel}
\end{equation}
We assumed a radius of $r_{\mathrm{max}} \sim 1\,\mathrm{Mm}$, and a peak rotational speed of $v_{\mathrm{max}} = 2.2\,\mathrm{km/s}$ ($\omega_0 = v_{\mathrm{max}}/r_{\mathrm{max}}$), typical for photospheric or chromospheric vortex motions observed in active regions \citep[e.g.,][]{Bonet2008, Galsgaard1996} and commonly used in numerical models of DC-type coronal heating (e.g., \citealt{howson2022effects, johnston2025self}).

Kink unstable coronal loop systems driven by photospheric twisting has been described in detail in previous studies \citep{reid2018coronal, cozzo2023coronal, cozzo2024coronal}. 
In the present work (as in \cite{2025A&A...695A..40C}), we focus instead on the subsequent phase of instability and address the scenario where a cold and dense filament is initially trapped inside the twisted flux rope.
We start with a hot ($1\,\mathrm{MK}$) and faint ($10^8\,\mathrm{cm}^{-3}$) corona and drive the magnetic twisting. Right before the flux tube reaches the threshold for kink instability, we introduce a condensation inside the twisted strand by multiplying (dividing) density $\rho_0$ (temperature $T_0$) by a Gaussian function with a minimum of 1 and a maximum of 100, $\sigma_x = \sigma_y = 1 \,\mathrm{Mm}$ and $\sigma_z = 10 \,\mathrm{Mm}$:
\begin{equation}
    \rho (t = 0) = \rho_0 \times \left(1 + 100\, e^{- \frac{x^2}{2 \sigma_x^2}} e^{- \frac{y^2}{2 \sigma_y^2}} e^{- \frac{z^2}{2 \sigma_z^2}} \right)
\end{equation}
\begin{equation}
    T (t = 0) = T_0 \div \left(1 + 100\, e^{- \frac{x^2}{2 \sigma_x^2}} e^{- \frac{y^2}{2 \sigma_y^2}} e^{- \frac{z^2}{2 \sigma_z^2}} \right)
\end{equation}
as shown in figure \ref{fig:initial_conditions}. The maximum total column density is $\sim 10^{20}\,\mathrm{cm}^2$, comparable with observations of prominences \citep{schwartz2015total}. We then let the kink instability evolve.

\subsection{Forward modelling}
\label{sec:forwardmodelling}

We synthesised the EUV intensity maps $I$ (DN~s$^{-1}$~pix$^{-1}$) in the SDO/AIA channels (94, 131, 171, 193, 211, 335, and 304~\AA), in the SolO/EUI 174~\AA\ band, and in the MUSE lines of \feix\ 171~\AA, \fexv\ 284~\AA, and \fexix\ 281~\AA\ by numerically integrating the emissivity ($j$) of each channel, weighted for their optical depth $\tau(x)$, through the 3D simulation box, along a selected line of sight:
\begin{equation}
    I (y,z) = \int_{\mathrm{LOS}} j(x) \,\exp[-\tau_i(x)]\,\mathrm{d}x,
\end{equation}
yielding synthetic maps directly comparable to EUV observations.

For each grid cell and channel, the optically-thin emissivity is evaluated as $j = n_e^2\,\Lambda(T)$, with $\Lambda(T)$ (DN~cm$^{5}$~s$^{-1}$~pix$^{-1}$) being the instrument emission response function vs plasma temperature and $n_e$ the free electron density. $\Lambda (T)$ combines the emission properties of the plasma with the sensibility of the instrument:
\begin{equation}
    \Lambda (T) = \int_0^{\infty} G(\lambda, T) \, R (\lambda) \, A_{\mathrm{pix}} \, d \lambda, 
\end{equation}
where $G(\lambda, T)$ is the plasma contribution function, $R (\lambda)$ is the spectral response, and $A_{\mathrm{pix}}$ is the pixel area. The temperature response functions are calculated using CHIANTI 10 \citep{del2021chianti} with the CHIANTI ionization equilibrium, coronal element abundances \citep{feldman1992elemental}, and a constant electron density of $10^9\,\mathrm{cm}^{-3}$.

To account for the eventual absorption by cool, partially ionized plasma, we considered the effect of photoionization extinction from neutral hydrogen \citep{labrosse2011euv} (the contributions from neutral helium and singly ionized helium have been neglected), using response-weighted effective cross sections $\tilde \sigma$ for each channel:
\begin{equation}
    \tilde{\sigma} \;=\;
    \frac{\displaystyle \int_{\lambda} R(\lambda)\,\sigma_{\rm HI}(\lambda)\,{\rm d}\lambda}
    {\displaystyle \int_{\lambda} R(\lambda)\,{\rm d}\lambda}\,
\end{equation}
Specifically, we evaluated the optical depth in front of each emitting element $(s, \infty)$ as:
\begin{equation}
    \tau(x)= \tilde \sigma \int_s^{\rm \infty} n_{\rm HI}(x')\,\mathrm{d}x'  
\end{equation}
with $n_{\rm HI}=f_{\rm HI}\,n_{\rm H}$ the density of neutral hydrogen (and $f_{\rm HI}$ its fraction \cite{Carlsson2012}). This constitutes only a minor correction to the fully optically thin approximation, as most hydrogen is fully ionized above $10^5,\mathrm{K}$ \citep{Carlsson2012}.

For MUSE spectroscopic observables, namely Doppler shifts and non-thermal line broadening, we computed, as a proxy, the first and second moments of the velocity distribution, weighted for the emissivity in each line (\citep{de2022probing, cozzo2024coronal, Cozzo_2026a}).

\subsection{DEM inversion}
\label{sec:deminversion}
The EM inversion from the intensity $\boldsymbol{I}$ of the six optically thin EUV channels of SDO/AIA (94, 131, 171, 193, 211, and 335~\AA), shown in Fig. \ref{fig:DEM_inversion_n1}, is obtained from the positive DEM distribution ($\boldsymbol{x}$) that solves the following minimisation problem:
\begin{equation}
\min_{\boldsymbol{x}\ge 0}\;\left\lVert \mathbf{W}\left(\mathbf{K}\boldsymbol{x}-\boldsymbol{I}\right)\right\rVert^{2}
+\lambda \left\lVert \mathbf{L}\boldsymbol{x}\right\rVert^{2},
\end{equation}
where: 
$\mathbf{K}$ is the discretised response matrix, $\mathbf{W}$ is a diagonal weighting matrix constructed from the intensity uncertainties, and $\mathbf{L}$ is a second-derivative operator enforcing smoothness. The non-negative solution was obtained via non-negative least squares (NNLS), while the regularisation strength $\lambda$ was selected using an L-curve criterion that balances data fidelity against solution smoothness \citep{hannah2012differential}. 

\subsection{Field lines tracing}
\label{sec:fieldlinestracing}
The field lines shown in Fig. \ref{fig:3D_rendering}, \ref{fig:3D_rendering_A1}, and \ref{fig:3D_rendering_A2} are computed using a fourth-order Runge-Kutta scheme, with the footpoints located in the negative-z side of the box, slowly moving according to the footpoint motions.

\subsection{ROAD algortithm}
\label{sec:roadalgorithm}
To automatically isolate the reconnection outflow shown in Fig. \ref{fig:horizontal_slice}, we used the ``ROAD" algorithm \citep{Cozzo_2026b}, which is based on the co-location of strong current sheets and high-speed flows. The method efficiently cluster slender, bi-directional structures with compact magnetic dissipation site (the current sheet) and (one or more) oppositely directed wings (the outflow).

\begin{appendices}

\section{Evolution of the kink instability}
\label{sec:kinkintability}

\begin{figure*}[t]
\centering
\includegraphics[width=\hsize]{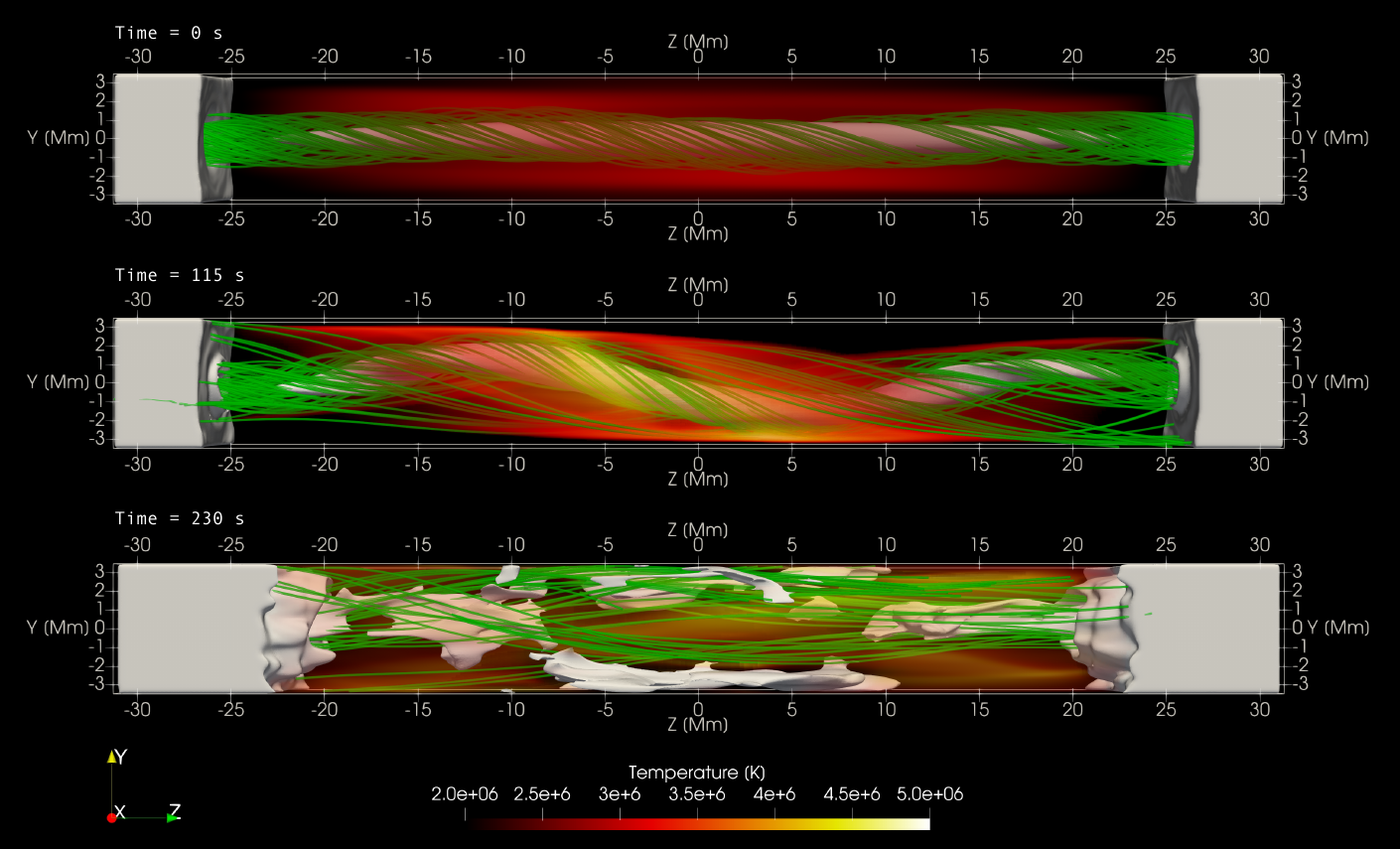}
\caption{3D rendering of the flux tube before (t = 0 s), during (t = 115 s) and after (t = 230 s) the kink instability. The volume rendering shows the high density \textcolor{black}{filament} regions ($\rho > 10^{10}\,\mathrm{cm}^{-3}$). Shades of red and yellow shows the coronal temperature distribution. Green magnetic field lines are traced from the left and right footpoints.}
\label{fig:kink_instability}
\end{figure*}

\begin{figure}[t]
\centering
\includegraphics[width=\hsize]{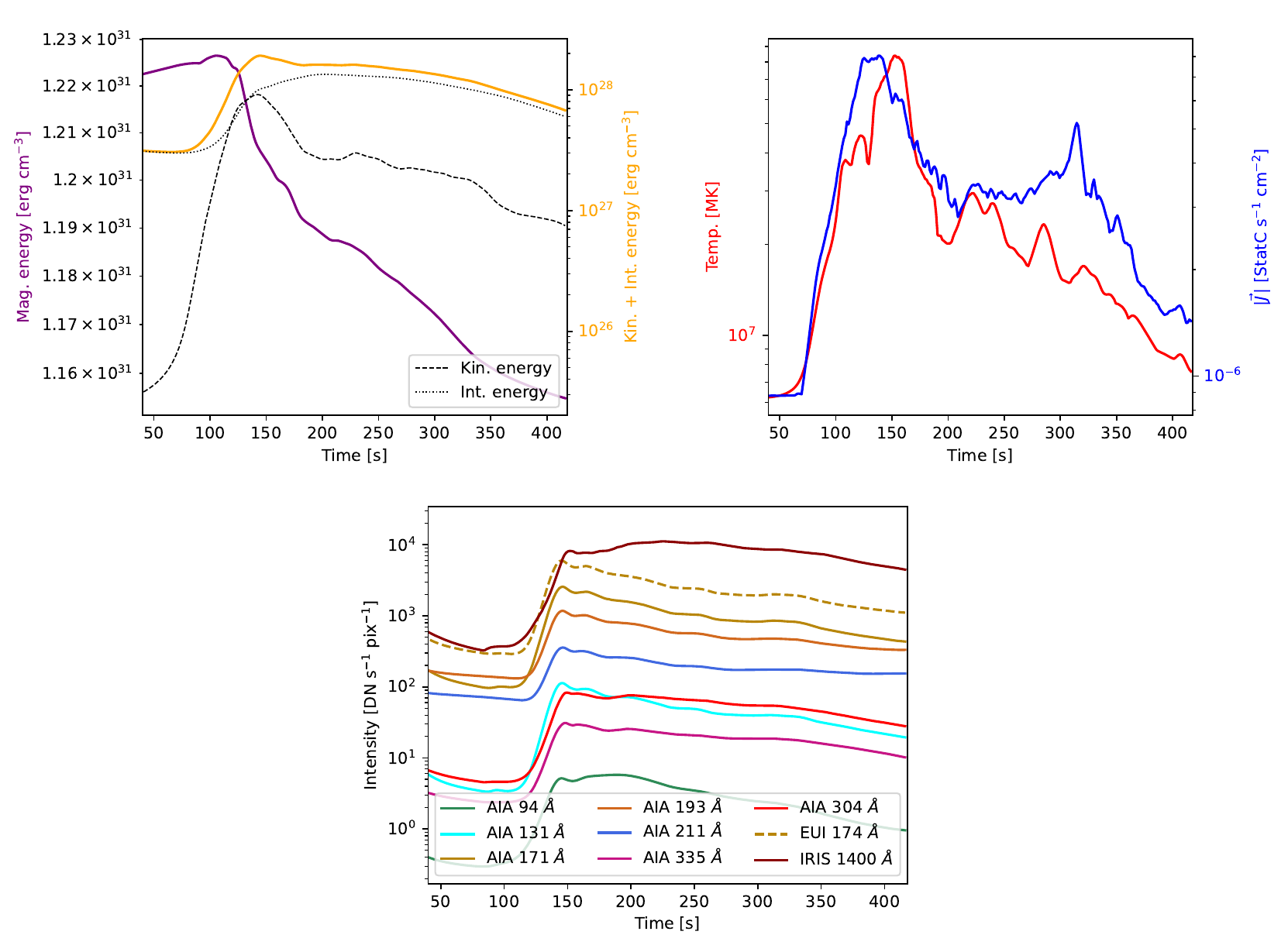}
\caption{Evolution of the total magnetic, kinetic, and internal energy (upper left), maximum temperature and current density (upper right), and averaged intensities (bottom) during the kink instability and later.}
\label{fig:1D_plot_kink}
\end{figure}

Figure \ref{fig:kink_instability} shows the initial kink instability before, during, and after the flux rope disruption. Initially ($t = 0\,\mathrm{s}$), the cold and dense material is trapped inside the twisted magnetic strand (Fig. \ref{fig:kink_instability}, upper panel). During the onset of the kink instability ($t \sim 100\,\mathrm{s}$), a helical perturbation develops and heats up the external shell of the flux tube (Fig. \ref{fig:kink_instability}, mid panel). At $t \sim 200\,\mathrm{s}$, the magnetic field structure and the material inside have fragmented (Fig. \ref{fig:kink_instability}, lower panel).
During the linear phase of the kink instability, the magnetic energy drops and releases about $10^{30}\,\mathrm{erg}$ into kinetic and internal energy (Fig. \ref{fig:1D_plot_kink}, left panel). The evolution is consistent with previous results \citep{hood2009coronal} that predict an exponential increase of the kinetic energy (with a secondary bump later on) and a smoother evolution of the internal energy.
In addition, the current density increases exponentially (Fig. \ref{fig:1D_plot_kink}, upper right panel), together with the temperature, as the helical current sheet develops \citep{hood2009coronal}. There is a short temporal delay between the current density peak and the temperature peak. Afterwards, the trajectory becomes more irregular because the magnetic field has fragmented into many current sheets, releasing localised, impulsive heating. 
After a short delay from the onset of the instability, also the intensities through the SDO/AIA, SolO/EUI, and IRIS channels follow an exponential trend, growing steeply by one order of magnitude in a few seconds (Fig. \ref{fig:1D_plot_kink}, bottom panel). \textcolor{black}{After the onset of the kink instability, the filament plasma fragments into multiple dense clumps, transitioning from a monolithic structure to a highly inhomogeneous medium with a morphology akin to coronal rain \citep{antolin2023extreme}, as the clumps subsequently fall along the magnetic flux tube under gravity.}

\backmatter

\backmatter

\section{Reconnection dynamics}
\label{sec:reconnectiondynamics}

\begin{figure*}[t]
\centering
\includegraphics[width=\hsize]{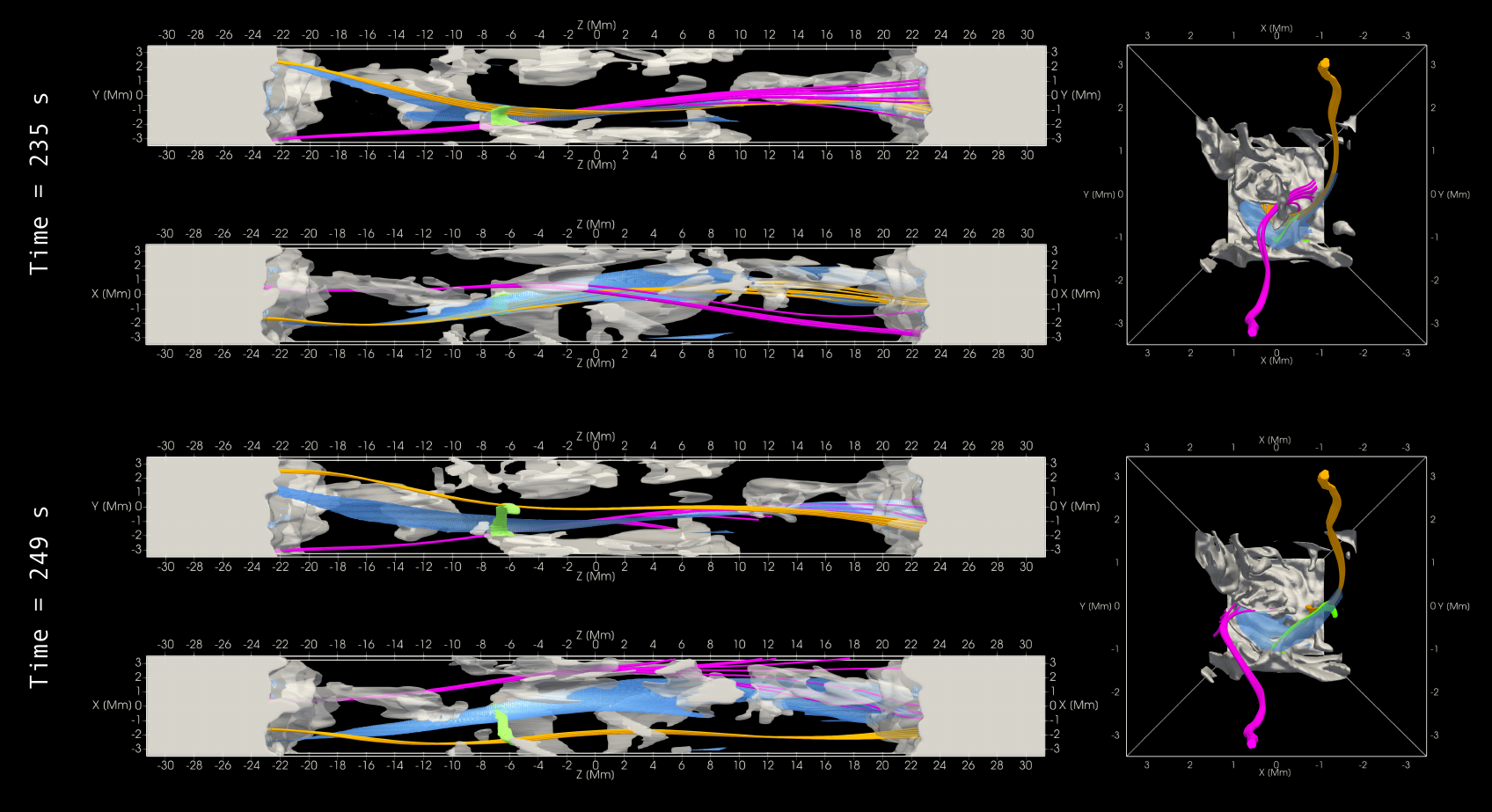}
\caption{3D rendering of the reconnection dynamics at the same times as in Fig. \ref{fig:3D_rendering}. For each snapshot time the three different orientations are shown. The white volume rendering shows the high density regions ($\rho > 10^{10}\,\mathrm{cm}^{-3}$). The nanojet plasma is emphasized by the green colour. Two reconnecting bundles of field lines are shown (magenta and orange), together with the current sheet (blue). A supplementary video is available.}
\label{fig:3D_rendering_A2}
\end{figure*}
\backmatter

The reconnection event that leads to the nanojet acceleration takes place around $z \sim 0\,\mathrm{Mm}$. This is emphasized in Fig. \ref{fig:3D_rendering_A2} by the orange and magenta field lines that cross each other around the center of the box at time $t = 235\,\mathrm{s}$ (upper panel). The nanojet originates from a region of dense plasma at $z \sim 7\,\mathrm{Mm}$ that is crossed by the orange line only. After $\sim 14\,\mathrm{s}$ (lower panel), the field lines have switched the magnetic connectivity at the (positive-z) footpoint, and the released magnetic tension has pushed them away from the current sheet. The lower panel \ref{fig:3D_rendering_A2} shows the orange line dragging part of the cold plasma, resulting in the detected nanojet (green volume). 

\begin{figure*}[t]
\centering
\includegraphics[width=\hsize]{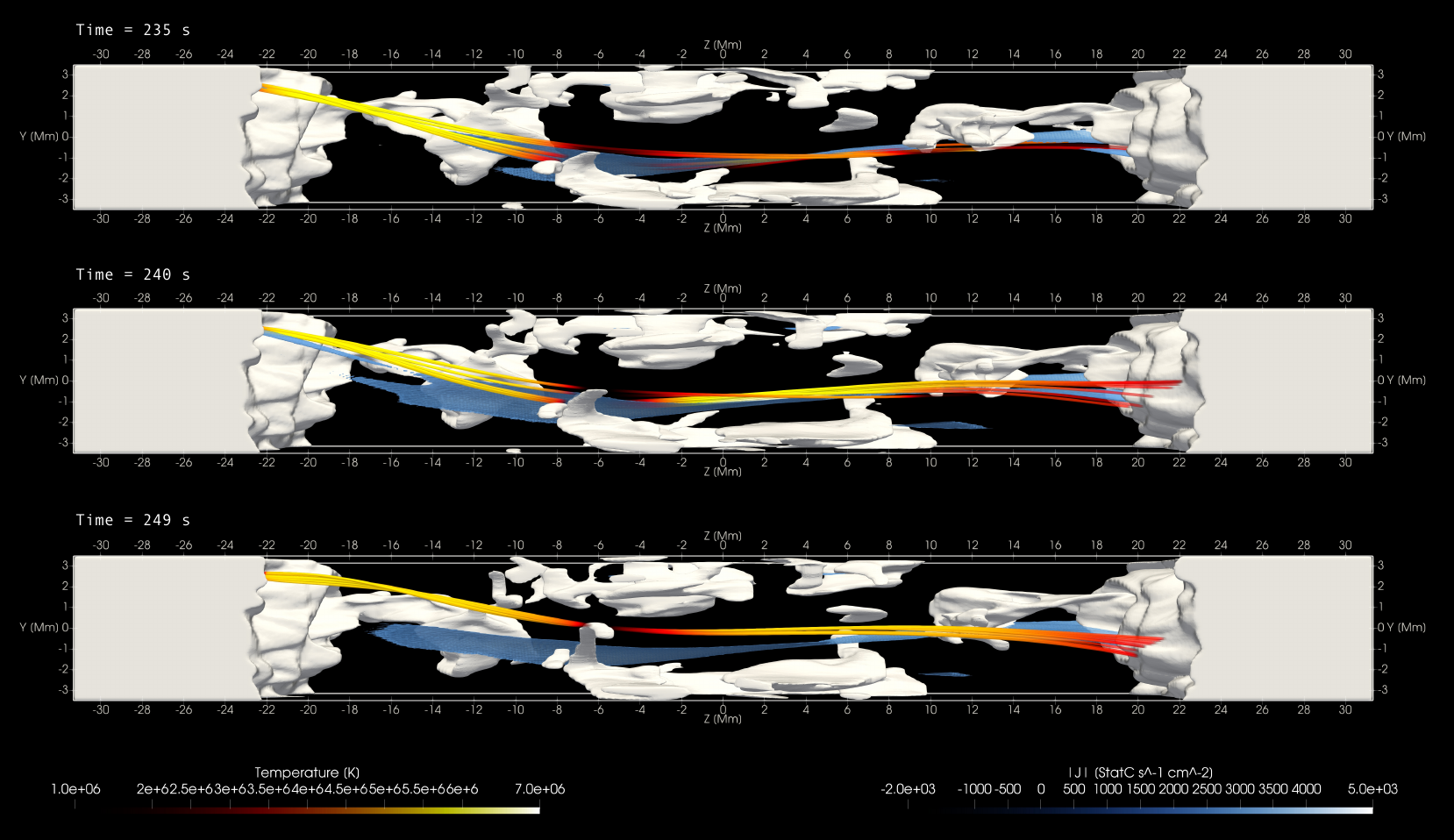}
\caption{3D rendering of plasma heating around the nanojet at three different times of its evolution. The white volume rendering shows the high density \textcolor{black}{filament} regions ($\rho > 10^{10}\,\mathrm{cm}^{-3}$). Magnetic field lines are colour-coded according to the plasma temperature. A current sheet passing through the jet is shown in blue. A supplementary video is available.}
\label{fig:3D_rendering_A1}
\end{figure*}

The reconnection process is accompanied by the heating of the tenuous plasma around the nanojet, as shown in Fig. \ref{fig:3D_rendering_A1}. The temperature of the plasma along the field lines is initially small ($\sim 2\,\mathrm{MK}$, upper panel) around $z \sim 0\,\mathrm{Mm}$, but, when the reconnection takes place (mid panel), it rapidly increases up to $6\,\mathrm{MK}$, and slowly decreases afterwards (lower panel).

\backmatter

\section{Analysis of synthetic observations}
\label{sec:timedistanceplots}

\begin{figure*}[t]
\centering
\includegraphics[width=\hsize]{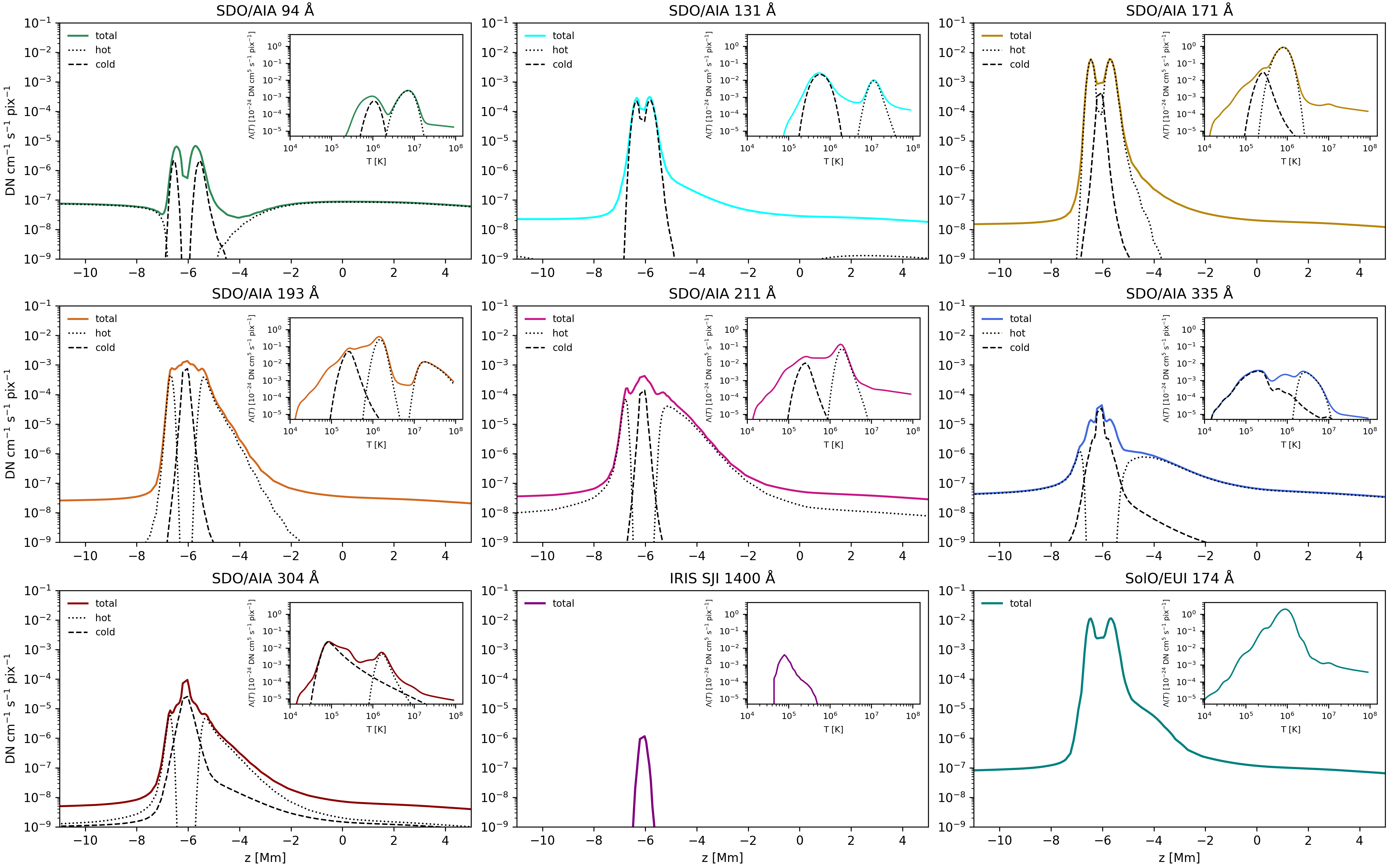}
\caption{Averaged emissivity in the SDO/AIA, SolO/EUI, and IRIS channels of Fig. \ref{fig:forward_modelling_n1}, along the field lines in Fig. \ref{fig:3D_rendering}, at time $t=249\,\mathrm{s}$. Total emission (solid lines), hot and cold contribution (dotted and dashed, respectively) are shown. The instrument response function for each channel is shown in the corresponding inset.}
\label{fig:emission_profile}
\end{figure*}

\begin{figure*}[t]
\centering
\includegraphics[width=\hsize]{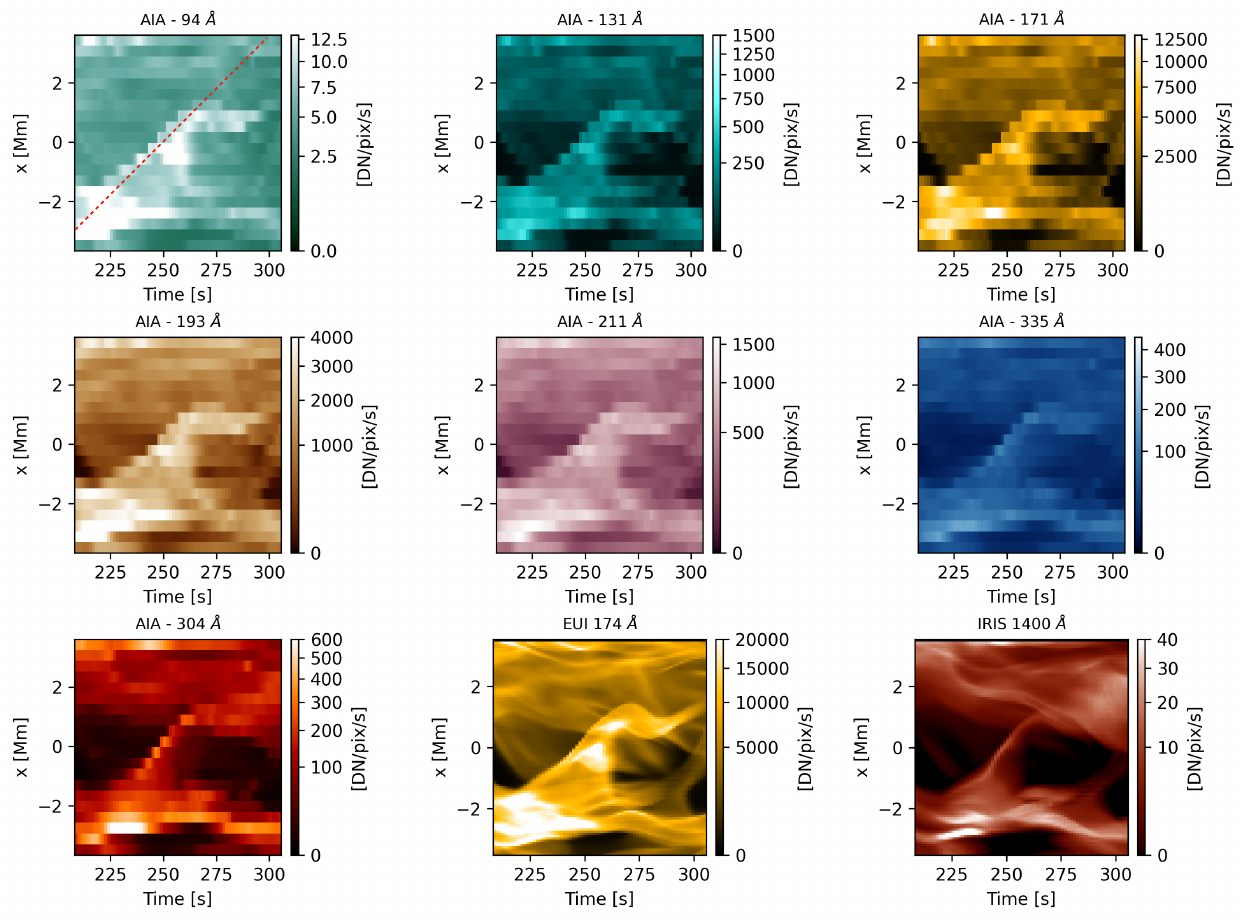}
\caption{Time distance maps of nanojet EUV emission along the white lines of figure \ref{fig:forward_modelling_n1}. The nanojet trajectory between $t = 230\,\mathrm{s}$, and $t = 260\,\mathrm{s}$ (dashed red line in the first panel) has slope of $70\,\mathrm{km}/\mathrm{s}$.}
\label{fig:time_distance_maps_1}
\end{figure*}

\begin{figure*}[t]
\centering
\includegraphics[width=\hsize]{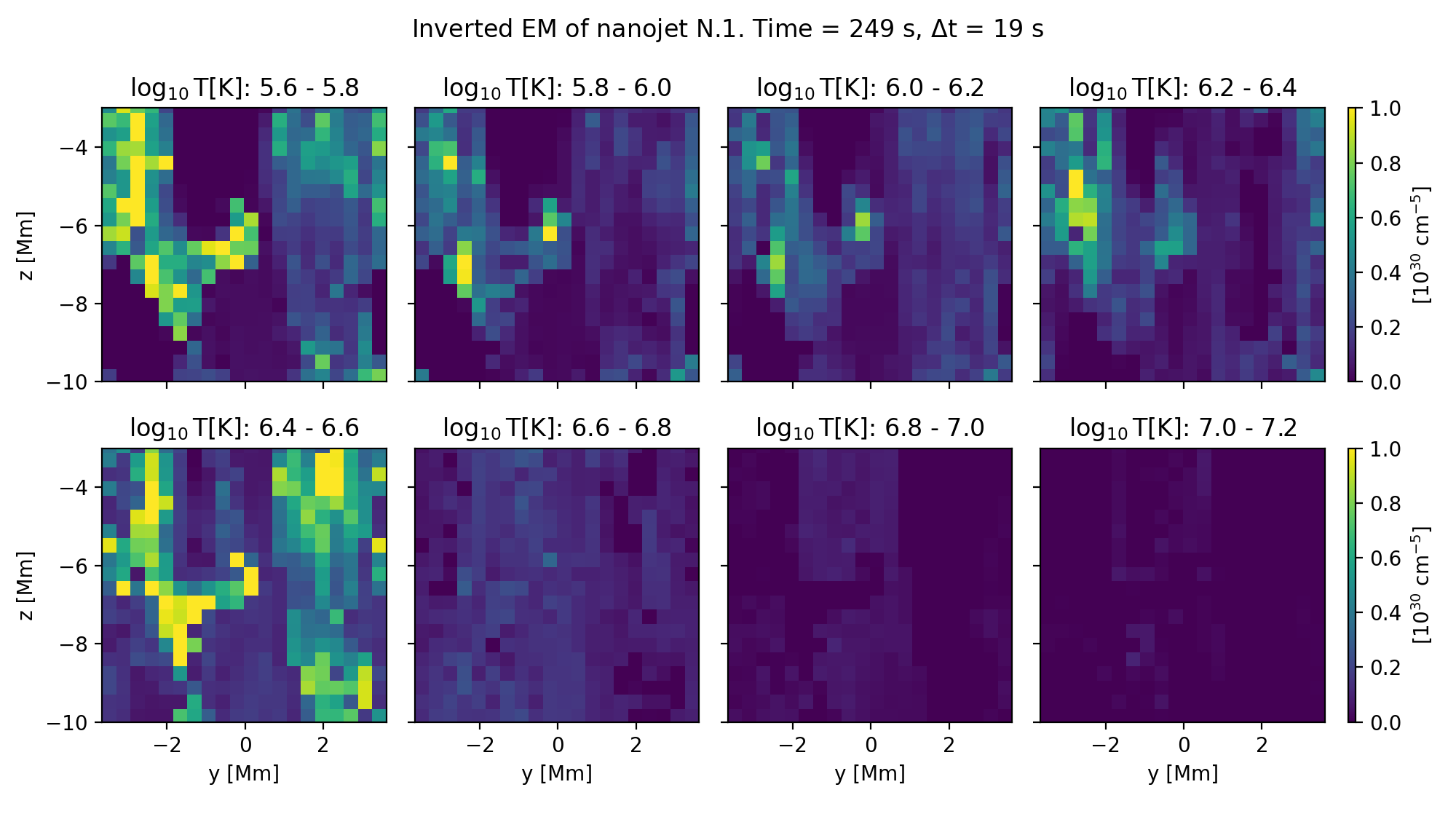}
\caption{Nanojet plasma emission measure (EM), as inverted from the AIA synthetic emission, across 8 logarithmically spaced temperature bins $\log_{10} T\,[\mathrm{K}] = 5.6$ to  $\log_{10} T\,[\mathrm{K}] = 7.2$. A supplementary video is available.}
\label{fig:DEM_inversion_n1}
\end{figure*}

\begin{figure*}[t]
\centering
\includegraphics[width=\hsize]{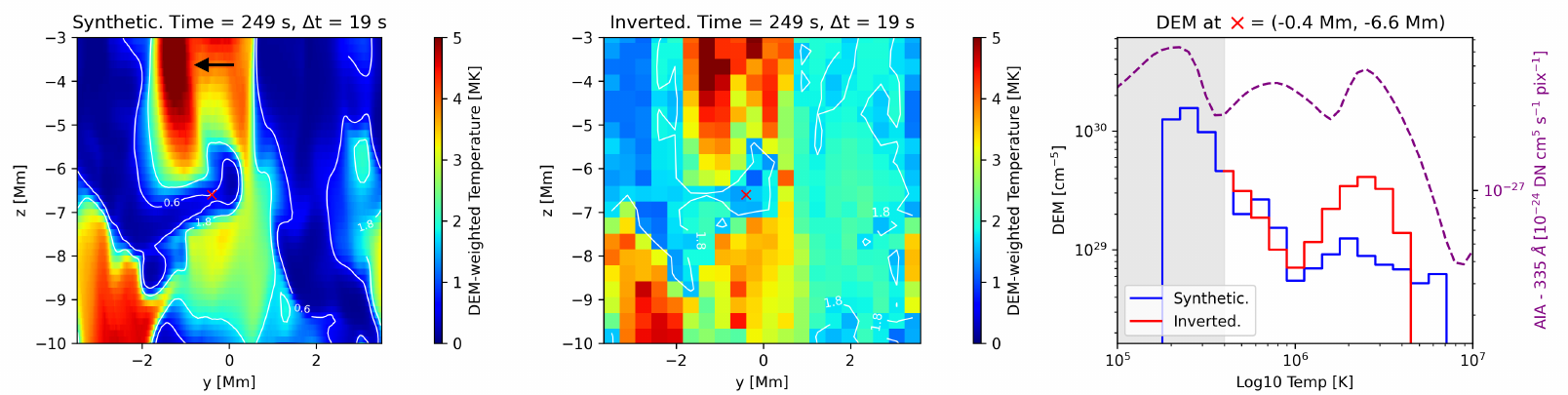}
\caption{Synthetic (first map) versus inverted (second map) DEM-weighted temperature of the region in Figs. \ref{fig:DEM_inversion_n1} and \ref{fig:DEM_synthesis_n1}. Isocontours at $T = 0.6\,\mathrm{MK}$ and $1.8\,\mathrm{MK}$ are plotted in white. The black arrow pinpoints very hot plasma in the vicinity of the reconnection point. Third panel: DEM distributions from the two methods at $\times =$(-0.4 Mm, -6.6 Mm), the grey area is outside the DEM-inversion temperature range. The AIA response function at 335 \AA\ is overplotted  (purple dashed line).}
\label{fig:DEM_weigthted_temperature_n1}
\end{figure*}
\backmatter

\begin{figure*}[t]
\centering
\includegraphics[width=\hsize]{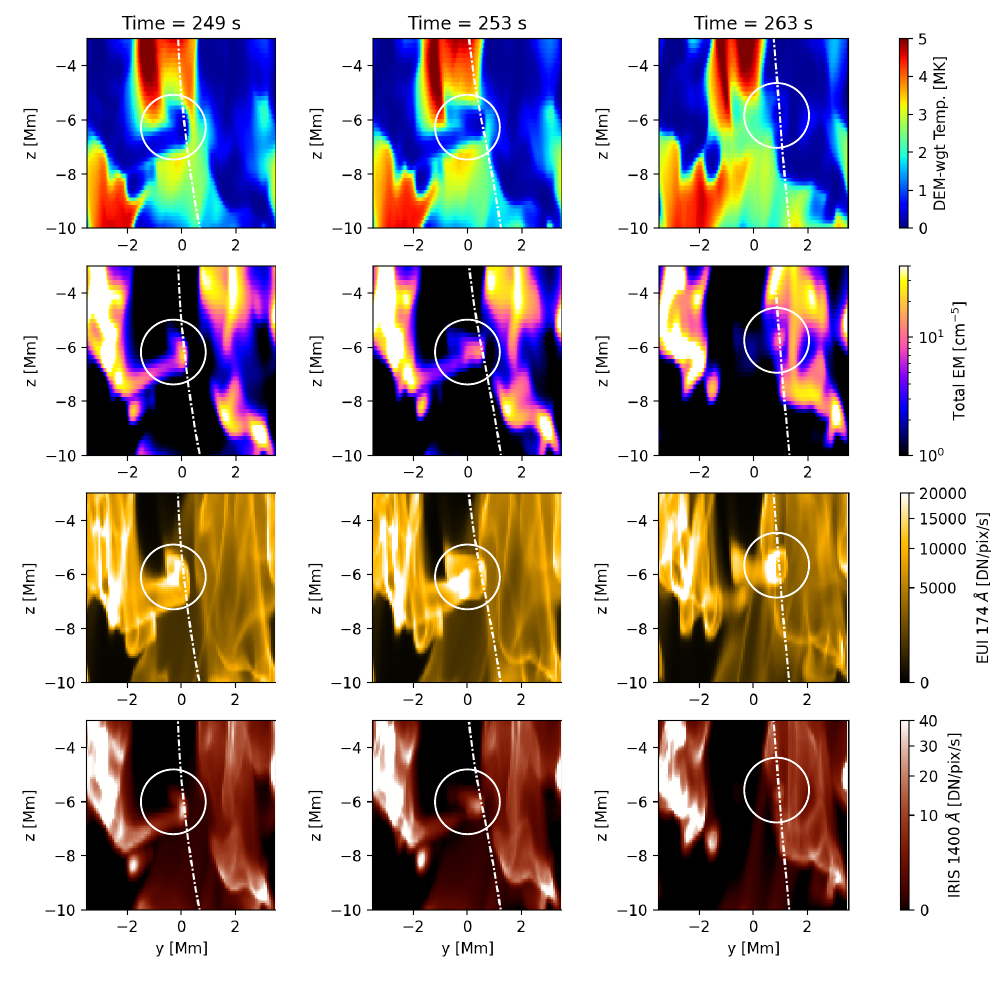}
\caption{Latest stages of the nanojet evolution. The panels show a zoom in of the temperature (first row), EM (second), and intensity in the EUI 174 \AA\ (third) and IRIS 1400 \AA\ (fourth) channels, at three different times towards the end of the nanojet journey. A (dot-dashed) field line is over-plotted. the moving plasmoid is circled.}
\label{fig:plasmoid}
\end{figure*}

To show in detail the respective contributions of cold and hot plasma to the intensity maps in Fig.~\ref{fig:forward_modelling_n1}, we plot in Fig.~\ref{fig:emission_profile} the emissivity profiles averaged over the set of field lines shown in Fig.~\ref{fig:3D_rendering}. For each AIA channel, we further decompose the signal into contributions from “cold” ($T\ll1\,\mathrm{MK}$) and “hot” ($T\gtrsim 1\,\mathrm{MK}$) plasma. In all channels, emission from hot plasma is either negligible or confined to the jet periphery, whereas the cold components exhibit a sharp peak at the nanojet core, indicating that the observed emission is primarily powered by dense ($\sim 10^{11}\,\mathrm{cm}^{-3}$) and cool ($\sim 10^5\,\mathrm{K}$) material.

The plane-of-sky nanojet speed, as inferred from its EUV emission, was estimated using time–distance maps shown in Fig. \ref{fig:time_distance_maps_1}. These were constructed by sampling the emission within the region bounded by the dashed lines in Fig.~\ref{fig:forward_modelling_n1} and, for each snapshot, averaging along the vertical direction. This procedure traces the nanojet trajectory in the $t$–$x$ plane. From the slope of this trajectory, we infer a projected in-plane velocity of $\approx 75\, \mathrm{km}\,\mathrm{s}^{-1}$.

The nanojet emission measure in Fig. \ref{fig:DEM_inversion_n1} was inferred from the synthetic maps in the six optically thin SDO/AIA EUV channels (94, 131, 171, 193, 211 and 335~\AA). We applied a regularized inversion method \citep{hannah2012differential} to recover the DEM from the multi-wavelength synthetic observations over the temperature range $\log_{10} T\,[\mathrm{K}] = 5.6$–$7.2$. We then compared the inverted EM to the “ground-truth” EM computed directly from the simulation data cube. The two distributions agree reasonably well at low temperatures (upper row in Fig. \ref{fig:DEM_inversion_n1} and Fig. \ref{fig:DEM_synthesis_n1}), but the inversion does not accurately reproduce the hot component (lower row). In particular, it produces an artificial excess in the $\log_{10} T\,[\mathrm{K}] = 6.4$–$6.6$ bin. Because several AIA temperature response functions are multi-peaked, this bias likely reflects a misattribution of emission in nominally high-temperature channels (e.g. 335~\AA) to genuinely hot plasma, when it can instead be produced by dense, cooler material.

The EM-weighted temperature in Fig.~\ref{fig:DEM_weigthted_temperature_n1} is computed from the synthetic (ground truth, left panel) and inverted (middle panel) emission measures shown in Figs.~\ref{fig:DEM_synthesis_n1} and \ref{fig:DEM_inversion_n1}. In the synthetic data, the nanojet contains a cold, narrow core at $\lesssim 0.5\,\mathrm{MK}$, surrounded by an envelope in which the temperature transitions from sub- to multi-MK values. In the inverted temperature maps, the nanojet temperature is generally overestimated, typically lying in the $1$–$2\,\mathrm{MK}$ range, with little evidence for sub-MK plasma. The right panel of Fig.~\ref{fig:DEM_weigthted_temperature_n1} shows the synthetic and inverted DEMs sampled at a representative nanojet location (marked by an “$\times$”). The overplotted AIA 335~\AA\ temperature response indicates that the spurious excess at $\sim 3\,\mathrm{MK}$ ($\log_{10}\,[\mathrm{K}] \sim 6.5$) arises from an overweighting of emission around the 335~\AA\ response peak.

Figure \ref{fig:plasmoid} focuses on the latest stages of the nanojet evolution. The initial dense and cold jet-like structure (left) rapidly detaches from the plasma condensation on the left (mid), with the head of the jet being transported by the moving field line. After $10\,\mathrm{s}$, it finally turns into a plasmoid (right), \textcolor{black}{as its jet-like structure fades and it transitions into a clump-like morphology.} It is still bright in the 174 \AA\ channel of EUI and slightly visible also with IRIS SJI at 1400 \AA. The plasmoid dissolves completely after a few tens of seconds.

\section{Spectroscopy with MUSE}
\label{sec:muse}

\begin{figure*}[t]
\centering
\includegraphics[width=\hsize]{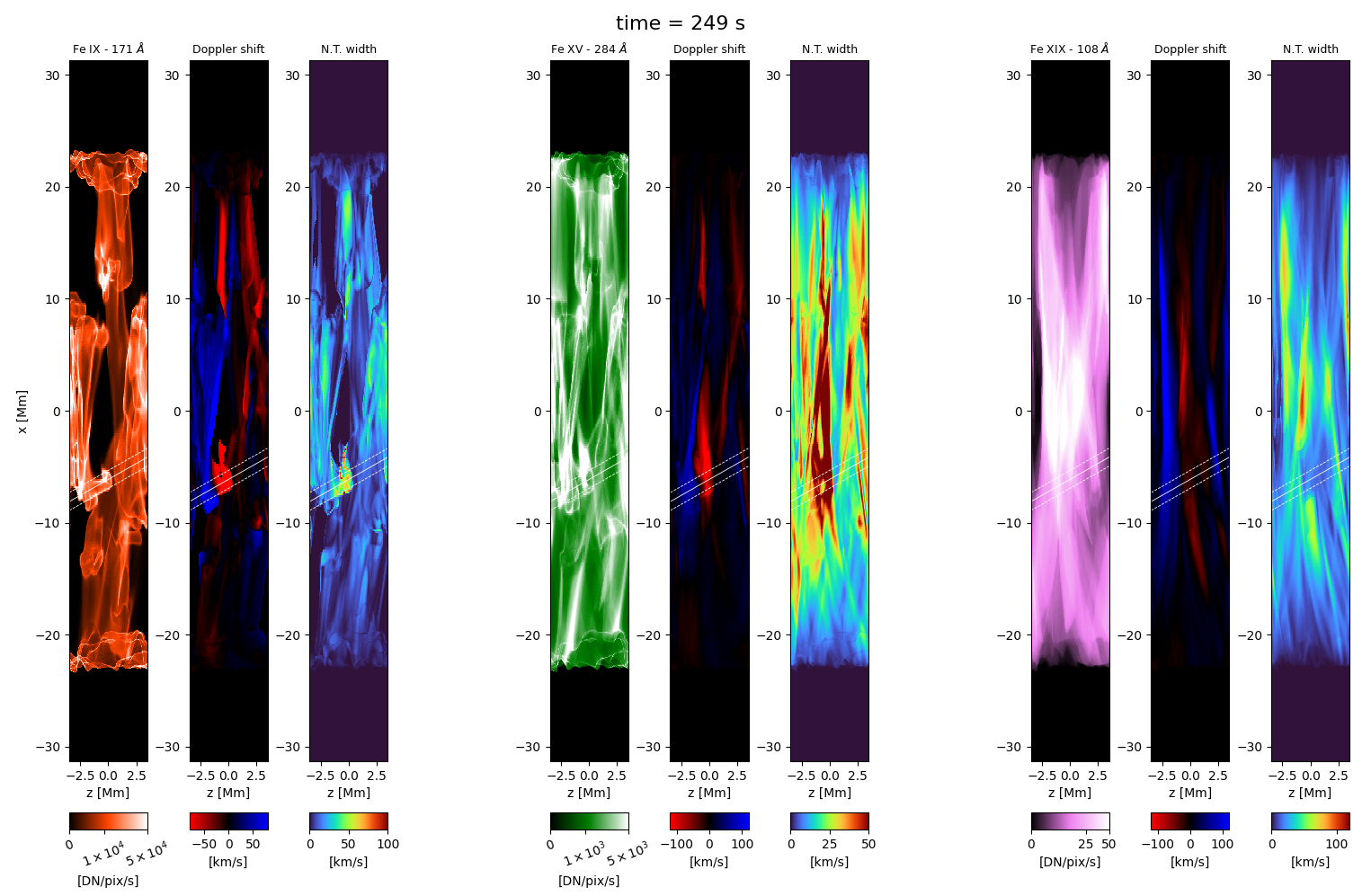}
\caption{Synthesis of the EUV emission, Doppler shifts and non-thermal line width across the three MUSE lines at 171 \AA\ (Fe IX), 284 \AA\ (Fe XV), and 108 \AA\ (Fe XIX). A solid white line aligns with the nanojet direction of propagation, dashed lines visually estimate the width. A supplementary video is available.}
\label{fig:MUSE_FM_n1}
\end{figure*}

\begin{figure*}[t]
\centering
\includegraphics[width=\hsize]{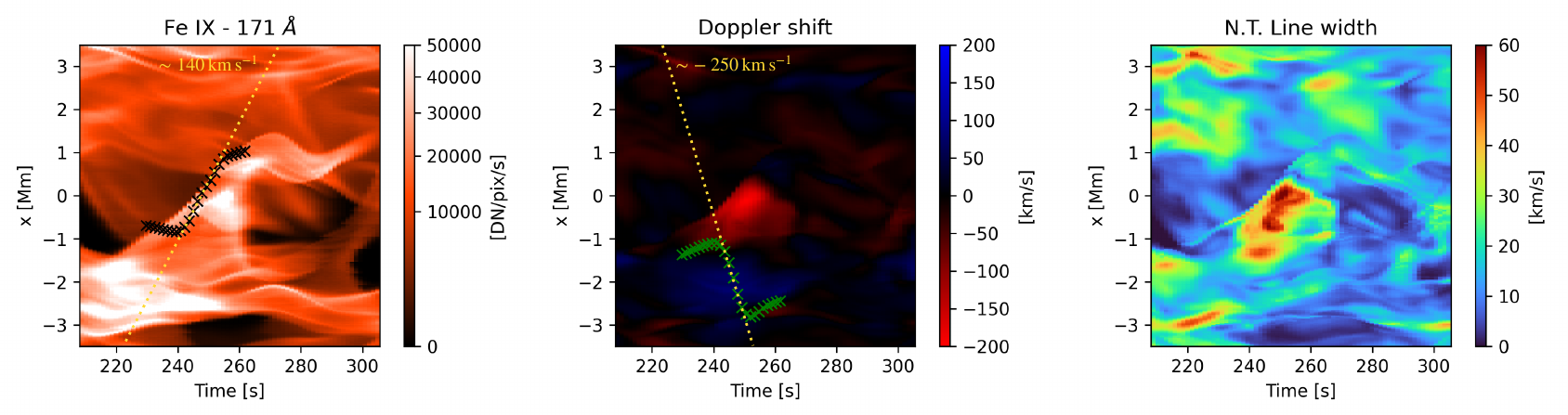}
\caption{Time distance maps of nanojet EUV emission, Doppler shifts, and non-thermal line width in the 171 \AA\ channel of MUSE along the white lines of figure \ref{fig:MUSE_FM_n1}. The ``$\times$" markers in the first (second) panel shows the averaged x (y) coordinate of field lines bundle shown in Fig. \ref{fig:3D_rendering} across the time. The $v_x$ ($v_y$) velocity component is estimated from the slope of dotted yellow line.}
\label{fig:MUSE_time_distance_maps_n1}
\end{figure*}

We forward-modelled the MUSE observables, namely the intensity of the three main lines (\feix\ at 171 \AA, \fexv\ at 284 \AA, and \fexix\ at 108 \AA), along with their respective Doppler shift and non-thermal broadening maps. The nanojet is visible only in the channel at 171 \AA, which is sensitive to lower temperatures ($\lesssim 1\,\mathrm{Mm}$). On the plane of the sky, as shown in Fig. \ref{fig:time_distance_maps_1}, the jet plasma travels at a speed of $\sim 70\,\mathrm{km}\,\mathrm{s}^{-1}$, which can be compared with the velocity of the moving field lines in Fig. \ref{fig:3D_rendering}. This latter velocity is estimated from the trajectory of the line in the time-distance plot of Fig. \ref{fig:MUSE_time_distance_maps_n1} (left panel) and is set at about $\sim 140\,\mathrm{km}\,\mathrm{s}^{-1}$. The head of the jet is red-shifted as it moves away from the observer. Along the LoS, the maximum plasma velocity registered by the Doppler map is $\sim 200\,\mathrm{km}\,\mathrm{s}^{-1}$, which is comparable with the projection of the field line velocity ($\sim 250\,\mathrm{km}\,\mathrm{s}^{-1}$, Fig. \ref{fig:MUSE_time_distance_maps_n1}, mid panel) but still smaller, as the partially diffusive plasma allows some slippage between the magnetic field and the plasma. A modest increase in the non-thermal line broadening is also registered (third panel on the right of Fig. \ref{fig:MUSE_FM_n1}, right panel in Fig. \ref{fig:MUSE_time_distance_maps_n1}).
The other two MUSE lines (\fexv\ at 284 \AA\ and \fexix\ at 108 \AA) do not show clear evidence of  a feature that resembles the nanojet, as they are sensitive to higher temperatures ($2.5\,\mathrm{MK}$ and $10\,\mathrm{MK}$, respectively). However, two long, blue- and red-shifted strands appear in the velocity maps of both channels, at the location of the reconnected field lines, linking the cold nanojet (bright at 171 \AA) with a hot reconnection outflow. Similarly, the non-thermal line broadening becomes strong along the reconnected magnetic field.

\backmatter

\backmatter

\section{Results for nanojet 2}
\label{sec:secondnanojet}

\begin{figure*}[t]
\centering
\includegraphics[width=\hsize]{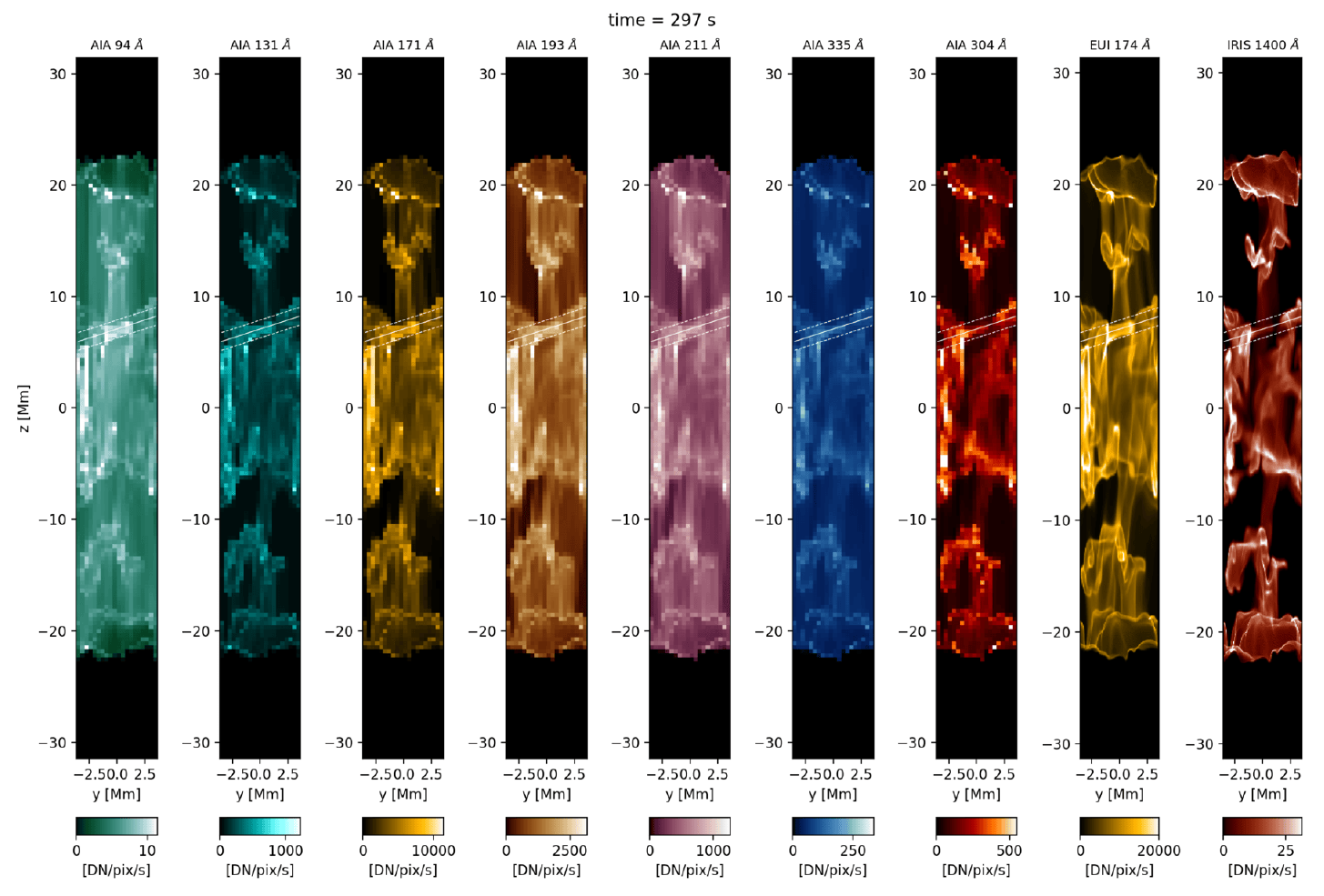}
\caption{Same as in Fig. \ref{fig:forward_modelling_n1} but for nanojet 2. A supplementary video is available.}
\label{fig:forward_modelling_n2}
\end{figure*}

\begin{figure*}[t]
\centering
\includegraphics[width=\hsize]{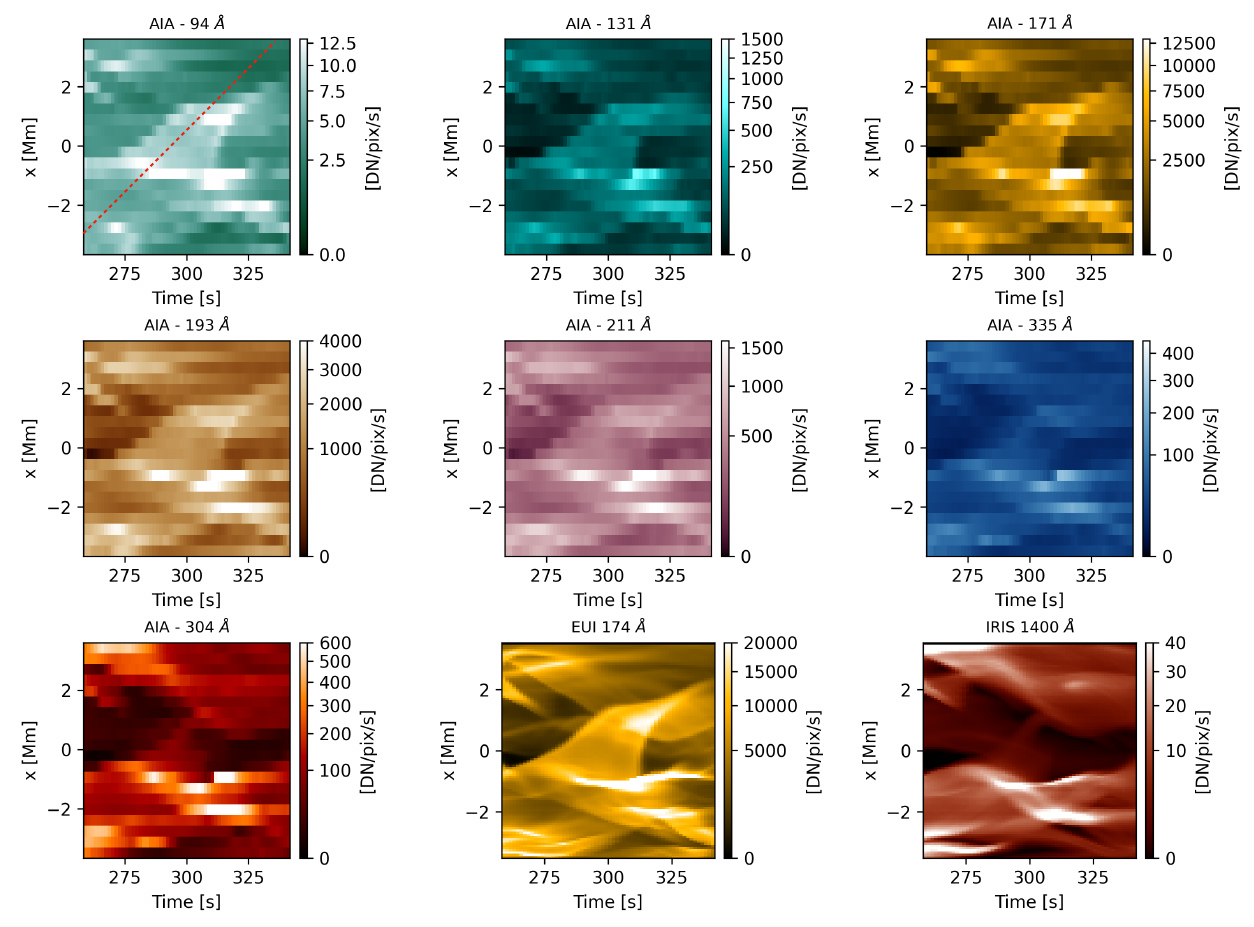}
\caption{Same as in Fig. \ref{fig:time_distance_maps_1} but for nanojet 2.}
\label{fig:time_distance_maps_n2}
\end{figure*}

\begin{figure*}[t]
\centering
\includegraphics[width=\hsize]{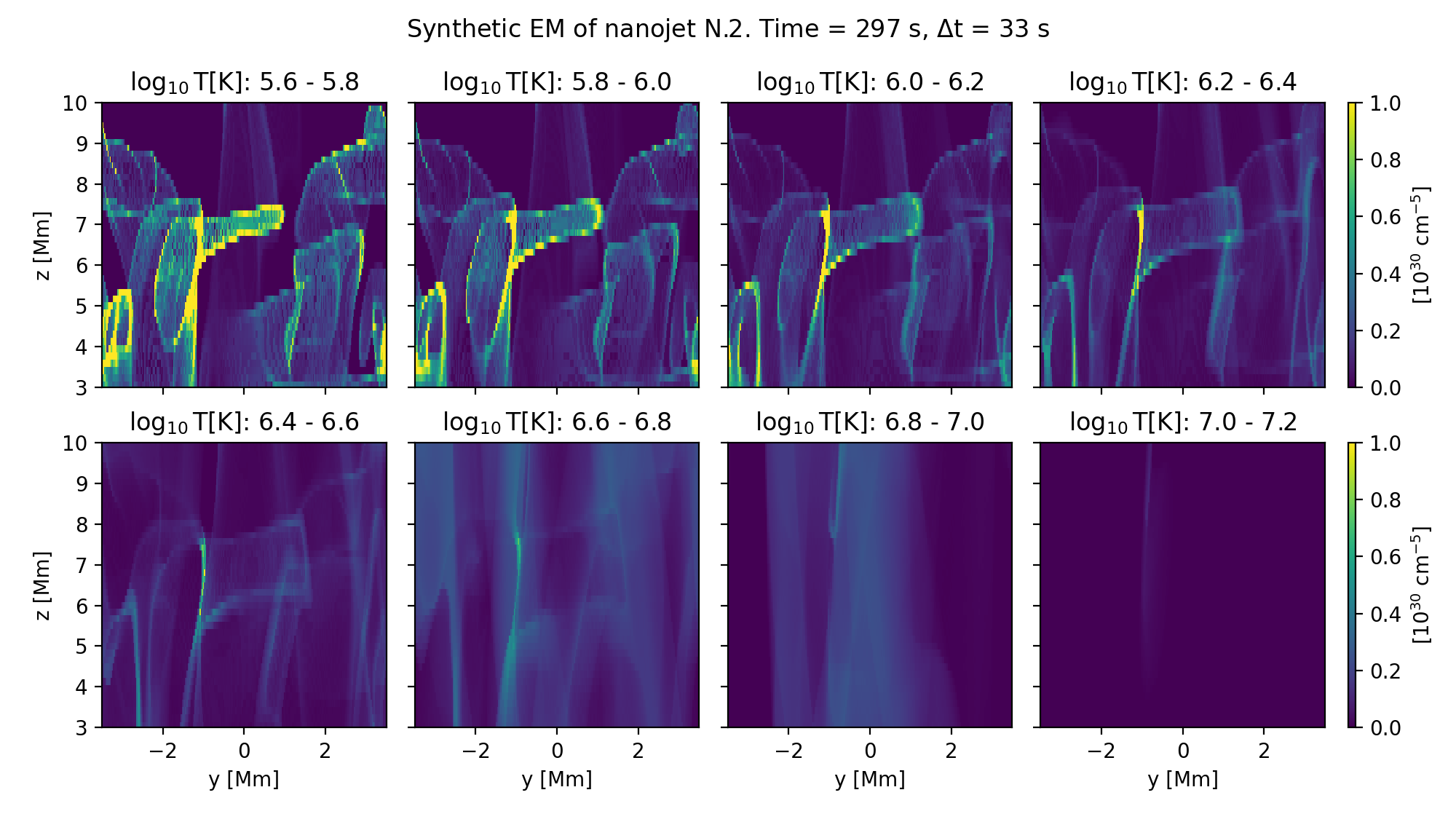}
\caption{Same as in Fig. \ref{fig:DEM_synthesis_n1} but for nanojet 2.}
\label{fig:DEM_synthesis_n2}
\end{figure*}

\begin{figure*}[t]
\centering
\includegraphics[width=\hsize]{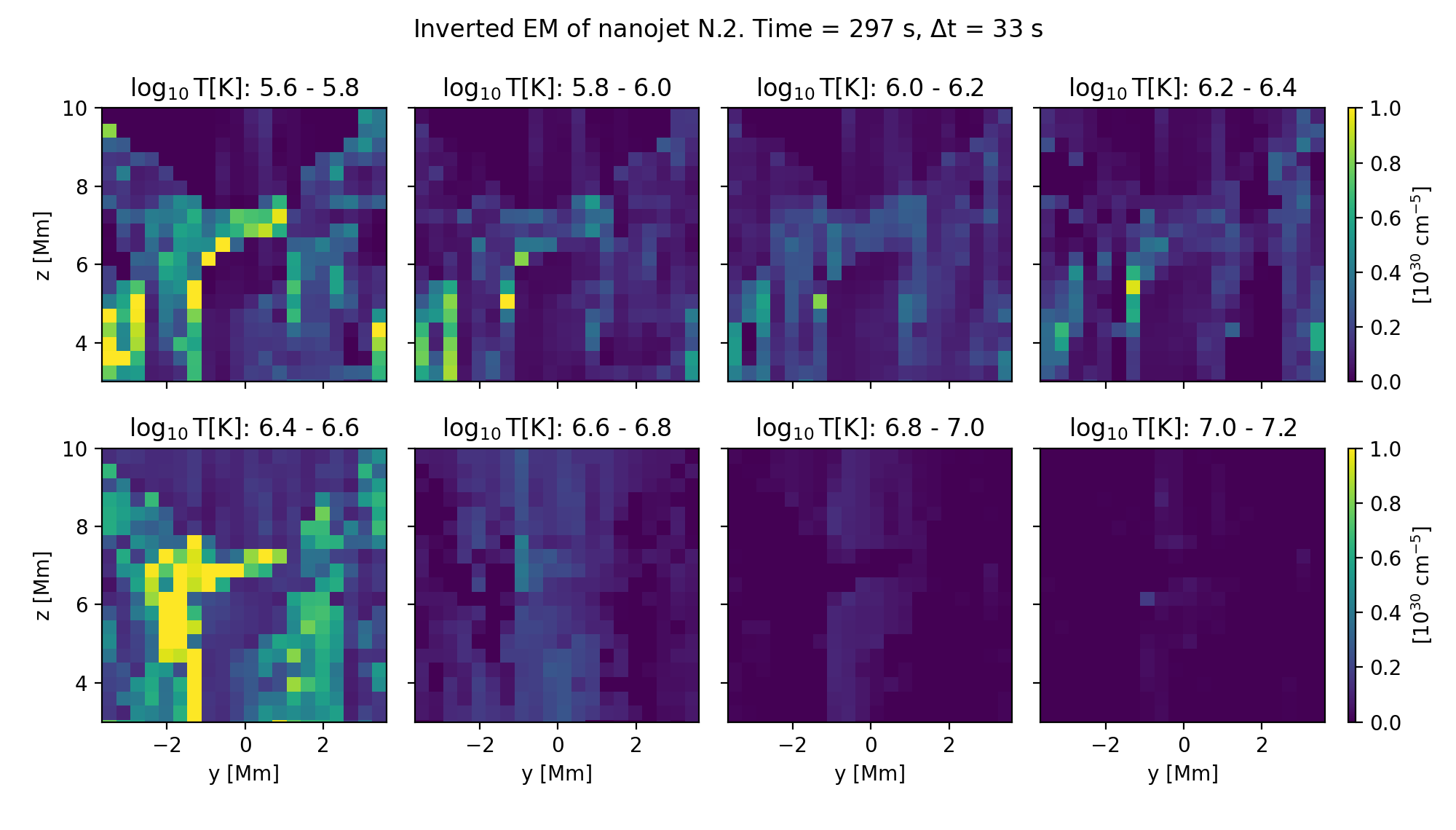}
\caption{Same as in Fig. \ref{fig:DEM_inversion_n1} but for nanojet 2. A supplementary video is available.}
\label{fig:DEM_inversion_n2}
\end{figure*}

\begin{figure*}[t]
\centering
\includegraphics[width=\hsize]{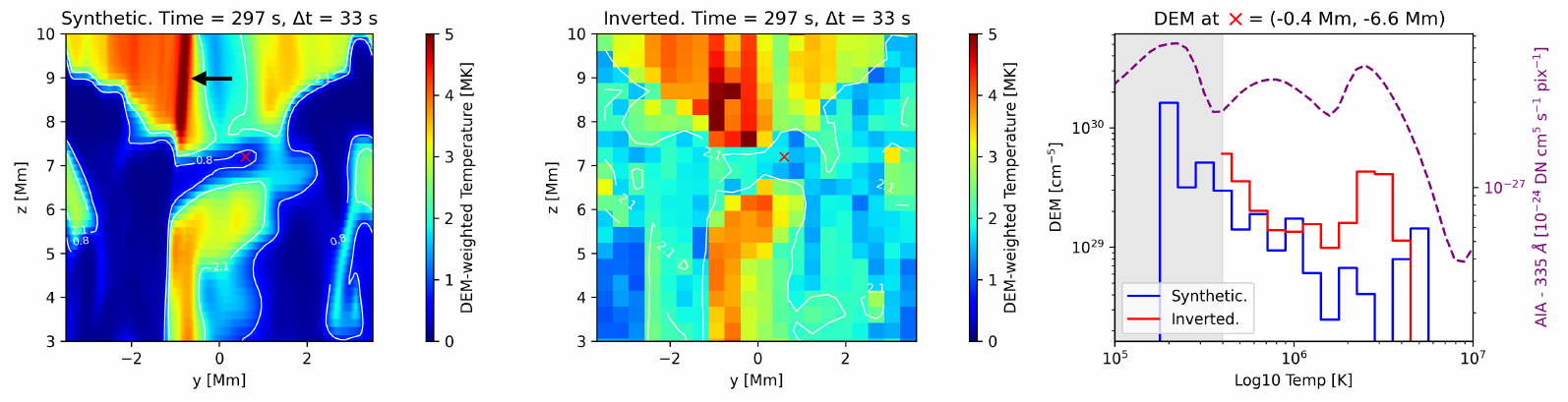}
\caption{Same as in Fig. \ref{fig:DEM_weigthted_temperature_n1} but for nanojet 2. Isocontours are at $T = 0.8$, and $2.1\,\mathrm{MK}$. Third panel plots are extracted from $\times =$(-0.4 Mm, 6.6 Mm). A supplementary video is available.}
\label{fig:DEM_weigthted_temperature_n2}
\end{figure*}

A second nanojet was detected from time $t \sim 275\,\mathrm{s}$ at $z \sim 7\,\mathrm{Mm}$. In the synthetic EUV images (Fig. \ref{fig:forward_modelling_n2}), it propagates for about $30\,\mathrm{s}$, reaching a maximum length of $\sim 2\,\mathrm{Mm}$ and an in-plane velocity of $\sim{75}\,\mathrm{km}\,\mathrm{s}^{-1}$, as inferred from the time-distance maps of Fig. \ref{fig:time_distance_maps_n2}. As for the first nanojet, it consists of cold and dense TR plasma (Fig. \ref{fig:DEM_synthesis_n2}) enveloped by increasingly hot coronal plasma. The DEM inversion in Fig. \ref{fig:DEM_inversion_n2} shows an excess of plasma in the $\log_{10}\,[\mathrm{K}]$:6.4-6.5 temperature bin, as the brightness in the AIA 335 \AA\ channel is incorrectly attributed to plasma at $\sim 3\,\mathrm{MK}$ (left panel of Fig. \ref{fig:DEM_weigthted_temperature_n2}), leading to an overestimate of the nanojet temperature (right v.s. mid panel of Fig. \ref{fig:DEM_weigthted_temperature_n2}).

From synthetic MUSE data, we can gather information about the nanojet LOS velocity and the dynamics of the hot coronal plasma around it. The head of the nanojet, visible in the 171 \AA\ channel, is blue-shifted at $\sim 50\,\mathrm{km}\,\mathrm{s}^{-1}$, giving a total velocity of $\sim 90\,\mathrm{km}\,\mathrm{s}^{-1}$.
The nanojet is no longer visible in the hot MUSE channels, but they help to pinpoint the region where reconnection takes place. Specifically, a narrow strand is bright in the 108 \AA\ channel and shows strong red-shifts and non-thermal broadening. It is located at the border of the cold plasma cloud from which the nanojet originates (arrow in Fig. \ref{fig:DEM_weigthted_temperature_n2}), suggesting that reconnection occurs nearby, rapidly heats up the plasma around it, and accelerates a reconnection outflow (from which the nanojet originates).

\backmatter

\backmatter

\begin{figure*}[t]
\centering
\includegraphics[width=\hsize]{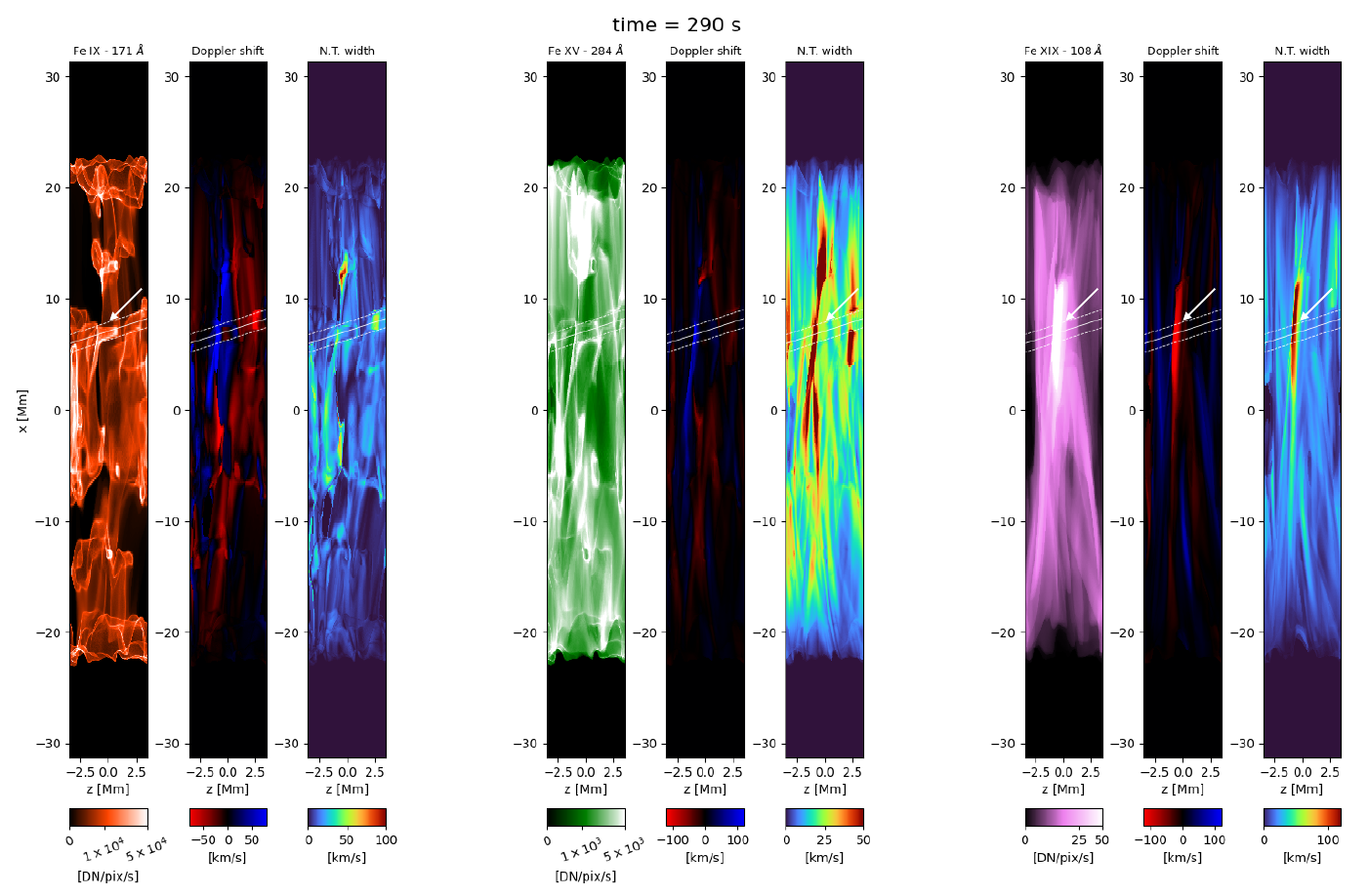}
\caption{Same as in Fig. \ref{fig:MUSE_FM_n1} but for nanojet 2. A supplementary video is available.}
\label{fig:MUSE_FM_n2}
\end{figure*}
\backmatter

\end{appendices}

\bibliography{sn-bibliography}

\end{document}